\pdfoutput=1
\documentclass[english,aps,nofootinbib,twocolumn,eqsecnum,reprint,superscriptaddress]{revtex4-1}
\usepackage[T1]{fontenc}
\usepackage[utf8]{inputenc}
\usepackage{amsmath}
\usepackage{amssymb}
\usepackage{graphicx}
\usepackage{setspace}
\usepackage{wasysym}
\usepackage{esint}

\makeatletter

\newcommand{\lyxdot}{.}

\usepackage{times}
\usepackage[utf8]{luainputenc}
\usepackage{fancyhdr}
\usepackage{array}
\usepackage{graphicx,amsmath,amsfonts,amssymb,slashed}
\usepackage{appendix}
\allowdisplaybreaks[4]

\usepackage{float}
\usepackage{wasysym}

\usepackage{esint}
\usepackage[marginal]{footmisc}
\usepackage{braket}
\usepackage{numprint}
\usepackage{enumitem}

\usepackage{verbatim}
\usepackage{numprint}
\usepackage{lipsum}
\usepackage{enumitem}
\usepackage{amsfonts}
\usepackage{mathrsfs}
\usepackage{scrextend}

\usepackage{setspace}

\usepackage{tikz}
\usepackage[colorlinks,linkcolor=blue,anchorcolor=blue,citecolor=blue,urlcolor=blue]{hyperref}

\usetikzlibrary{chains,patterns,scopes,positioning,backgrounds,shapes,fit,shadows,calc,arrows.meta,decorations.pathreplacing,calligraphy}

\renewcommand{\baselinestretch}{1.1}
\makeatother

\makeatother

\usepackage{babel}
\begin{document}
\title{Dynamic solar Primakoff process}
\author{Zheng-Liang Liang}
\email{liangzl@mail.buct.edu.cn}

\affiliation{College of Mathematics and Physics, Beijing University of Chemical
Technology~\\
Beijing 100029, China}
\author{Lin Zhang}
\email{zhanglin57@mail.sysu.edu.cn}

\affiliation{School of Science, Shenzhen Campus of Sun Yat-sen University~\\
 Shenzhen, 518107, China}
\affiliation{Sun Yat-sen University~\\
 Guangzhou, 510275, China}
\begin{abstract}
The Primakoff mechanism is one of the primary channels for the production
of solar axion. In canonical estimation of the Primakoff photon-axion
conversion rate, the recoil effect is neglected and a static structure
factor is adopted. By use of the linear response theory, we provide
a dynamic description of the solar Primakoff process. It is found
that the collective electrons overtake ions as the dominant factor,
in contrast to the static screening picture where ions contribute
more to the photon-axion conversion. Nonetheless, the resulting axion
flux is only 1\textasciitilde 2\% lower than the standard estimate
based on the static structure factor.
\end{abstract}
\maketitle

\section{Introduction}

The QCD axion that emerged originally as a solution to the strong-\textit{CP}
problem~\citep{Peccei:1977hh,Weinberg:1977ma,Wilczek:1977pj}, is
also well motivated as a promising dark matter (DM) candidate~\citep{Preskill:1982cy,Abbott:1982af,Dine:1982ah,Cadamuro:2011fd}
other than the weakly interacting massive particles (WIMPs), and has
attracted increased interests in both theoretical and experimental
fronts in recent years. The rich phenomenology of axion can leave
peculiar traces in cosmology, astroparticle physics and particle physics~\citep{Raffelt:1985nk,Raffelt:1990yz,Raffelt:1996wa,Duffy:2009ig,Kawasaki:2013ae,Marsh:2015xka,Graham:2015ouw}.

The Sun is the primary natural source for terrestrial axion detection.
With a coupling to standard model (SM) particles, axions can be produced
in the solar interior through a number of channels, such as the Primakoff
process~\citep{Primakoff:1951iae,Dicus:1978fp} and the axio-recombination,
bremsstrahlung and Compton scattering (ABC) process~\citep{Dimopoulos:1986kc,Redondo:2013wwa}.
For Kim-Shifman-Vainshtein-Zhakharov (KSVZ) axions~\citep{DINE1981199,Zhitnitsky:1980tq},
the former reaction dominates, while for the Dine-Fischler-Srednicki-Zhinitsky
(DFSZ) axions~\citep{Kim:1979if,Shifman:1979if}, the latter mechanism
dominates. In this paper, we focus on the Primakoff axion production
mechanism, where photons convert to axions through the Coulomb field
sourced by the charged particles (i.e., electrons and ions).

In conventional wisdom~\citep{Raffelt:1985nk}, the charged particles
in the Sun are so heavy compared to the energies of ambient photons
that they can be regarded as fixed, in which case the photon energy
in a scattering event is considered equal to that of the emitted axion,
and the differential cross section of the Primakoff process $\mathrm{d}\sigma_{\gamma\rightarrow a}\left(\mathbf{p}_{\gamma}\right)/\mathrm{d}\varOmega$
is proportional to $\left|\mathbf{p}_{\gamma}\times\mathbf{p}_{a}\right|^{2}/Q^{4}$,
with $\mathbf{p}_{\gamma}$ and $\mathbf{p}_{a}$ being the momenta
of the incident photon and emitted axion, respectively, and $Q=\left|\mathbf{Q}\right|=\left|\mathbf{p}_{\gamma}-\mathbf{p}_{a}\right|$.
In the massless limit of an axion, the cross section is divergent
due to the long-range Coulomb interaction. This Coulomb potential
can be regulated if the solar in-medium screening effect is taken
into account. Raffelt~\citep{Raffelt:1985nk} argued that the implication
of screening effect on the differential cross section is described
with the substitution,
\begin{eqnarray}
\frac{\left|\mathbf{p}_{\gamma}\times\mathbf{p}_{a}\right|^{2}}{Q^{4}} & \rightarrow & \frac{\left|\mathbf{p}_{\gamma}\times\mathbf{p}_{a}\right|^{2}}{Q^{4}}S\left(\mathbf{Q}\right)\nonumber \\
 & = & \frac{\left|\mathbf{p}_{\gamma}\times\mathbf{p}_{a}\right|^{2}}{Q^{4}}\frac{1}{1+\kappa^{2}/Q^{2}},\label{eq:raffelt}
\end{eqnarray}
where the Debye-Hückel scale $\kappa$ can be as large as $\sim9\,\mathrm{keV}$
at the solar center, and effectively provides a cutoff of the Coulomb
interaction. Note that this description is based on the assumption
of a negligible recoil effect in the solar medium, and thus a static
structure factor $S\left(\mathbf{Q}\right)=\left(1+\kappa^{2}/Q^{2}\right)^{-1}$
is introduced to measure the correlation between the charged particle
density~\citep{Raffelt:1985nk,Raffelt:1996wa}.

In Ref.~\citep{Raffelt:1987np}, Raffelt further considered finite
energy shifts in the axion production process in the presence of classical
electric-field fluctuations in the solar plasma, so that the collective
electron motion and the its implication for the axion conversion are
unified under a general framework. Moreover, by use of the Kramers-Kronig
relations that relate the spectral densities in the electromagnetic
fluctuation description to the static structure factor $S\left(\mathbf{Q}\right)$,
Raffelt reasoned that the Primakoff production rate in Ref.~\citep{Raffelt:1985nk}
agrees with the total rates of the decay process $\gamma_{t}\,\left(\mathrm{transverse\,plasmon}\right)\rightarrow\gamma_{l}\,\left(\mathrm{longitudinal\,plasmon}\right)+a\,\left(\mathrm{axion}\right)$,
the plasma coalescence process $\gamma_{t}+\gamma_{l}\rightarrow a$,
and the individual Primakoff process $\gamma_{t}+e/N\,\left(\mathrm{electron/ion}\right)\rightarrow a+e/N$.
Thus, as far as the calculation of the axion production rate is concerned,
the static structure factor $S\left(\mathbf{Q}\right)$ has already
included the collective behavior of the solar medium. These two seemingly
different descriptions reflect the same electromagnetic properties
of the solar medium.

Refs.~\citep{Altherr:1990tf,Altherr:1992jg,Altherr:1993zd} reproduced
the same Primakoff production rate using the thermal field theory,
by the same Kramers-Kronig relations argument in the last step. It
should be noted that in order for the Kramers-Kronig sum rules to
work, the energy shift between the axion and the photon $\omega$
is assumed to be remarkably smaller than the solar temperature $T_{\odot}$
such that $1-e^{-\omega/T_{\odot}}\simeq\omega/T_{\odot}$. Thus a
detailed numerical examination of this assumption is one of the inspirations
for this work.

Recently, in Ref.~\citep{Liang:2023ira} we applied the nonrelativistic
linear response theory to the dynamic screening effect in the nondegenerate
gas of the solar plasma associated with the dark matter scattering.
Under this framework, one no longer needs to add the dielectric function
by hand, since both the finite temperature effect and the many-body
effect are inherently encapsulated in the dynamic structure factor
$S\left(\mathbf{Q},\omega\right)$. ``Dynamic'' means that a finite
energy $\omega$ transfer and thus a temporal variation is taken into
account in a scattering event, in contrast to the static case where
the charged particles are regarded as fixed targets. This is important
considering the thermal velocities of electrons can reach $\sim0.1\,c$
in the core of the Sun, which may bring a non-negligible Doppler energy
shift in the Primakoff process. And more importantly, the screening
and the collective effect (plasmon) are naturally incorporated into
this dynamic structure factor $S\left(\mathbf{Q},\omega\right)$.
This method could be an alternative to the electromagnetic fluctuation
description~\citep{Raffelt:1987np} and the thermal field theory
approach~\citep{Altherr:1990tf,Altherr:1992jg,Altherr:1993zd} mentioned
above.

Therefore, the purpose of this work is to apply the linear response
theory approach~\citep{Liang:2023ira} to the photon-axion conversion
process inside the Sun, in order to investigate the implication of
the recoil effect and the collective effect, and especially to numerically
explore in detail to what extent the Kramers-Kronig sum rule argument
is reliable to validate the calculation of the Primakoff conversion
rate based on the static structure factor.

Discussion will proceed in the natural units, where $\hbar=c=k_{B}=1$.

\section{\label{sec:EELS}Primakoff event rate}

We first introduce how we describe the Primakoff process in the context
of the linear response theory that naturally encodes the relevant
finite temperature physics and the many-body in-medium effect.
\begin{figure}[b]
\begin{centering}
\includegraphics[scale=0.35]{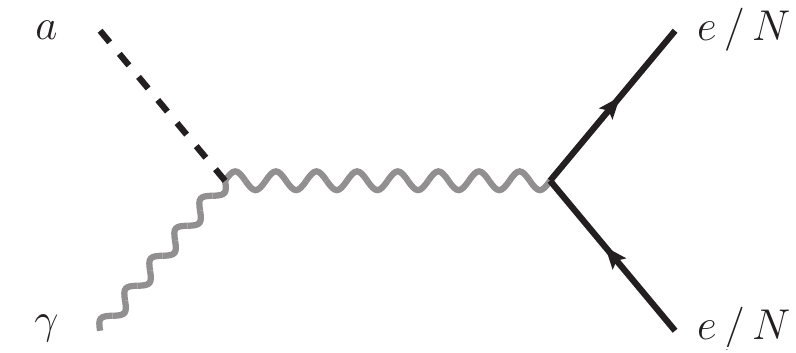}
\par\end{centering}
\caption{\textit{\label{fig:Primakoff}}Diagram for the Primakoff scattering
process where a photon is converted into an axion in the Coulomb potential
of charged particles (electron and ions). See the text for details.}
\end{figure}

At the effective field theory (EFT) level, the interaction relevant
for the Primakoff process is given as
\begin{eqnarray}
\mathcal{L}_{a\gamma} & = & -\frac{g_{a\gamma}}{4}aF_{\mu\nu}\tilde{F}^{\mu\nu},\label{eq:Lagrangian}
\end{eqnarray}
where $a$ is the axion field, $g_{a\gamma}$ represents the axion-photon
coupling, and $F_{\mu\nu}$ and $\tilde{F}_{\mu\nu}=\frac{1}{2}\epsilon_{\mu\nu\rho\sigma}F^{\rho\sigma}$
are the electromagnetic field strength and its dual, respectively.
\begin{figure}[b]
\begin{centering}
\includegraphics[scale=0.77]{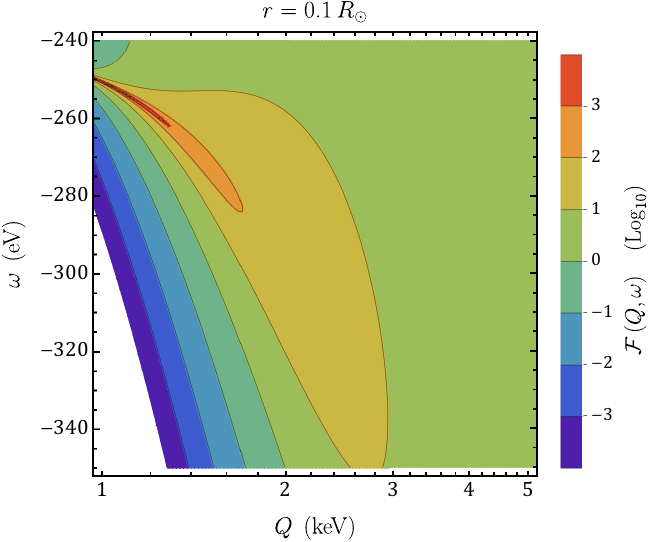}
\par\end{centering}
\caption{\label{fig:PrimakoffSpectrum} The factor $\mathcal{F}\left(Q,\omega\right)$
that demonstrates the longitudinal plasmon resonance of the solar
medium at radius $r=0.1\,R_{\odot}$ (with $R_{\odot}$ being the
solar radius). The highly resonant “line” (below $Q<1\,\mathrm{keV}$)
is too narrow to be shown in the plot. It is evident that such plasmon
suffers strong damping for $Q\apprge2\,\mathrm{keV}$. See the text
for details.}
\end{figure}

Since the electrons and ions move nonrelativistically in the Sun,
we express the relevant interactions in the nonrelativistic EFT. For
instance, the electromagnetic field-electron interaction is written
as
\begin{eqnarray}
\mathcal{L}_{Ae} & = & -eA_{0}\psi_{e}^{*}\psi_{e}-\frac{ie}{2m_{e}}\mathbf{A}\cdot\left(\psi_{e}^{*}\overrightarrow{\nabla}\psi_{e}-\psi_{e}^{*}\overleftarrow{\nabla}\psi_{e}\right)\nonumber \\
 &  & -\frac{e^{2}}{2m_{e}}\left|\mathbf{A}\right|^{2}\cdot\psi_{e}^{*}\psi_{e}+\cdots,
\end{eqnarray}
where $\psi_{e}$ is the nonrelativistic electron wavefunction. Considering
that the second term on the right-hand side is subject to an electron
velocity suppression $\frac{\nabla}{m_{e}}\sim v_{e}\sim\mathcal{O}\left(10^{-2}\sim10^{-1}\right)$
in the solar medium~(with $m_{e}$ being the electron mass), and
the longitudinal and transverse photon propagators do not mix under
the random phase approximation (RPA), only the longitudinal component
in the nonrelativistic effective electron-photon interaction $A_{0}$~(or
more specifically, the Coulomb interaction) is retained for the description
of electron-electron~(and electron-ion) interaction in this work.
Thus, we only consider the components $-g_{a\gamma}a\,\epsilon^{ijk0}\partial_{i}A_{j}\partial_{k}A_{0}$
of the Lagrangian in Eq.~(\ref{eq:Lagrangian}) in the estimate of
the Primakoff process in the Sun. While the $A_{0}$ component is
responsible for the Coulomb interaction, $\left\{ A_{j}\right\} $
are relevant for the transverse photon external legs.

The calculation of the Primakoff process shown in Fig.~\ref{fig:Primakoff}
depends on an accurate description of the electronic and ionic in-medium
effect inside the Sun. In this work, we invoke the linear response
approach proposed in Ref.~\citep{Liang:2023ira} to describe the
screening effect in the Sun. Within this framework, the axion production
rate for an incident photon with energy $E_{\gamma}$ and momentum
$\mathbf{p}_{\gamma}$ can be summarized by the following expression
(see the Appendix~\ref{sec:AppendixA} for further details): \begin{widetext}
\begin{eqnarray}
\Gamma\left(\mathbf{p}_{\gamma}\right) & = & \int\mathrm{d}\omega\int\frac{\mathrm{d}^{3}Q}{\left(2\pi\right)^{3}}\frac{g_{a\gamma}^{2}\,\left|\mathbf{p}_{\gamma}\times\mathbf{Q}\right|^{2}}{8\,E_{\gamma}\,\sqrt{\left|\mathbf{p}_{\gamma}-\mathbf{Q}\right|^{2}+m_{a}^{2}}}\frac{1}{Q^{2}}\,\delta\left(\sqrt{\left|\mathbf{p}_{\gamma}-\mathbf{Q}\right|^{2}+m_{a}^{2}}-E_{\gamma}+\omega\right)\nonumber \\
 &  & \times\frac{\left(-2\right)}{1-e^{-\omega/T_{\odot}}}\,\left[\frac{V_{e}\,\mathrm{Im}\left(\Pi_{e}\right)}{\left|1-V_{e}\,\Pi_{e}-V_{e}\,{\displaystyle \sum\limits _{i}Z_{i}^{2}\,\Pi_{N_{i}}}\right|^{2}}+\frac{V_{e}\,\sum\limits _{i}Z_{i}^{2}\,\mathrm{Im}\left(\Pi_{N_{i}}\right)}{\left|1-V_{e}\,\Pi_{e}-V_{e}\,{\displaystyle \sum\limits _{i}Z_{i}^{2}\,\Pi_{N_{i}}}\right|^{2}}\right],\label{eq:Primakoff rate}
\end{eqnarray}
\end{widetext}where $\alpha=e^{2}/4\pi$ is the electromagnetic fine
structure constant, $\omega$ and $Q=\left|\mathbf{\mathbf{Q}}\right|$
denote the energy and the magnitude of the momentum transfer to the
solar medium, respectively, $V_{e}\left(Q\right)=4\pi\alpha/Q^{2}$
is the electron Coulomb interaction in momentum space, and $E_{a}$
($m_{a}$) is the energy (mass) of the axion. The delta function represents
the energy conservation in the scattering.

In this work, we only consider the case where the axion masses are
so small~(typically $\ll$ keV) compared to their energies that they
can be effectively considered as massless. Besides, Eq.~(\ref{eq:Primakoff rate})
does not take into account the fact that photons inside a plasma have
a nontrivial dispersion relation $E_{\gamma}^{2}\simeq\omega_{p}^{2}+\left|\mathbf{p}_{\gamma}\right|^{2}$,
which means photons propagating in the solar medium have an effective
mass $\omega_{p}=\sqrt{4\pi\alpha n_{e}/m_{e}}$. Consequently, the
conversion process is only possible for $E_{\gamma}>\omega_{p}$,
and the plasmon mass effect becomes remarkable for energies $E_{\gamma}\gtrsim\omega_{p}$.
However, since $\omega_{p}\approx0.3\,\mathrm{keV}$ is much smaller
than the typical plasmon energy $3T_{\odot}\approx4\,\mathrm{keV}$
in the solar core~\citep{Vogel2023}, where the majority of axions
are produced, the photons are also treated as massless in Eq.~(\ref{eq:Primakoff rate}),
in line with the treatment in Ref.~\citep{Raffelt:1985nk}. Ref.~\citep{Hoof:2021mld}
provided an analytical conversion rate that adds the plasma frequency
to the original expression in Ref.~\citep{Raffelt:1985nk}, based
on which we verify that the plasmon mass effect brings a correction
at the order of $10^{-4}$ between 1 and $12\,\mathrm{keV}$.

The first and second terms in the square brackets in Eq.~(\ref{eq:Primakoff rate})
correspond to the finite temperature many-body effect from the electrons
and ions, respectively; while $\mathrm{Im}\left(\Pi_{e}\right)$ ($\mathrm{Im}\left(\Pi_{N_{i}}\right)$)
is responsible for the thermal movement of the electrons (ions), the
denominator describes the screening~\citep{Liang:2023ira}. $\Pi_{e}$
denotes the electron one-particle-irreducible diagram, in the RPA
which is approximated as the bubble diagram. For the nondegenerate
electron gas in the Sun, $\Pi_{e}$ can be expressed as~\citep{Liang:2023ira}
\begin{widetext}
\begin{eqnarray}
\Pi_{e}\left(Q,\,\omega\right) & = & -\frac{n_{e}}{Q}\sqrt{\frac{m_{e}}{2\,T_{\odot}}}\left\{ \Phi\left[\sqrt{\frac{m_{e}}{2\,T_{\odot}}}\left(\frac{\omega}{Q}+\frac{Q}{2\,m_{e}}\right)\right]-\Phi\left[\sqrt{\frac{m_{e}}{2\,T_{\odot}}}\left(\frac{\omega}{Q}-\frac{Q}{2\,m_{e}}\right)\right]\right\} \nonumber \\
\nonumber \\
 &  & -i\,n_{e}\sqrt{\frac{2\pi}{m_{e}T_{\odot}}}\left(\frac{m_{e}}{Q}\right)\exp\left[-\left(\frac{m_{e}^{2}\,\omega^{2}}{Q^{2}}+\frac{Q^{2}}{4}\right)\frac{1}{2\,m_{e}T_{\odot}}\right]\sinh\left[\frac{\omega}{2\,T_{\odot}}\right],\label{eq:polarizability}
\end{eqnarray}
\end{widetext}where $n_{e}$ is the number density of the electron
gas, and the function $\Phi$ is defined as the Cauchy principal value
of the integration~\citep{fetter2012quantum}
\begin{eqnarray}
\Phi\left(x\right) & \equiv & \mathcal{P}\int_{-\infty}^{+\infty}\frac{\mathrm{d}y}{\sqrt{\pi}}\frac{e^{-y^{2}}}{x-y}.
\end{eqnarray}

\begin{figure*}
\begin{centering}
\includegraphics[scale=0.75]{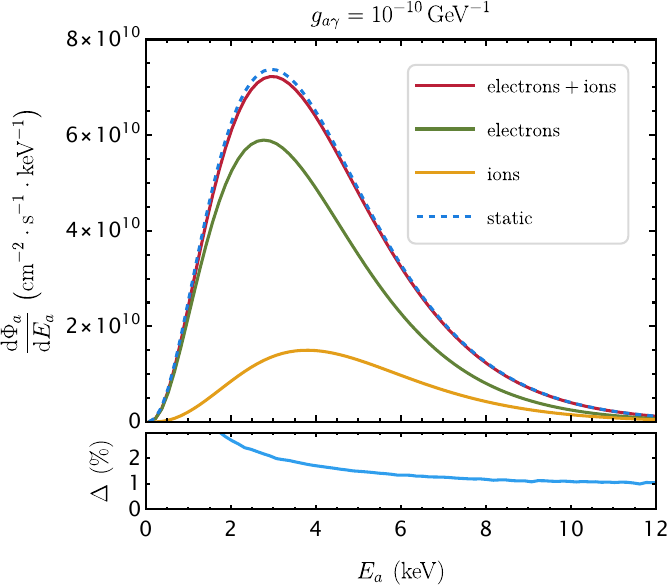}\hspace{0.5cm}\includegraphics[scale=0.78]{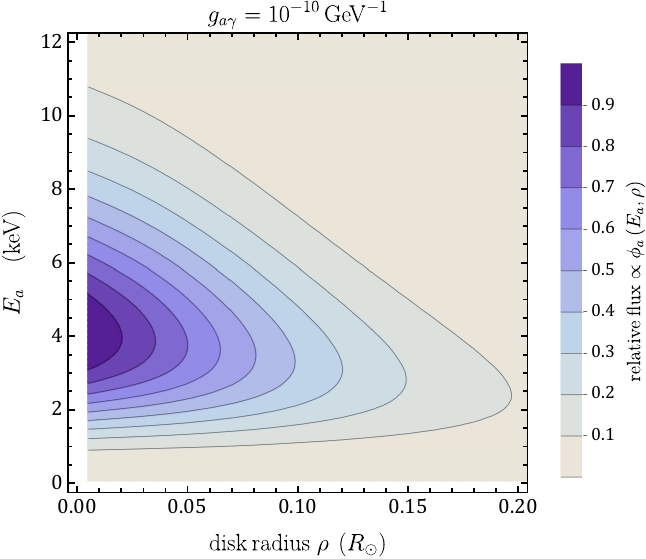}
\par\end{centering}
\caption{\label{fig:ElectronSpectrum} \textit{Left}: Solar Primakoff axion
spectra on Earth calculated with the static structure factor (blue
dashed), and with the dynamic structure factor (red) that consists
of the contributions from electrons (green) and ions (orange), for
a benchmark coupling strength $g_{a\gamma}=10^{-10}\mathrm{GeV}^{-1}$.
$\Delta$ (in $\%$) represents the relative difference between the
two approaches, i.e., $\Delta\equiv\left(\mathrm{static}-\mathrm{dynamic}\right)/\mathrm{static}$.
\textit{Right}: The contour plot of the solar axion surface luminosity,
normalized to its maximum value, depending on the radius $\rho$ on
the solar disk and energy $E_{a}$. See the text for details.}
\end{figure*}
Similarly, $\Pi_{N_{i}}$ denotes the bubble diagram of $i^{\mathrm{th}}$
ion species carrying a charge $Z_{i}e$. At the RPA level, $\Pi_{N_{i}}$
can be obtained by simply replacing $n_{e}$ and $m_{e}$ with ion
number density $n_{N_{i}}$ and ion mass $m_{N_{i}}$ in Eq.~(\ref{eq:polarizability}).
Contributions from all solar ion species are included in Eq.~(\ref{eq:Primakoff rate}).

In order to describe the collective behavior of the solar medium,
we introduce a nondimensional function $\mathcal{F}\left(Q,\omega\right)$,
which represents the second line in Eq.~(\ref{eq:Primakoff rate}).
Interestingly, from Fig.~\ref{fig:PrimakoffSpectrum} a strong resonance
structure is observed in the parameter area where the real part approaches
zero in the denominator in Eq.~(\ref{eq:Primakoff rate}), which
corresponds to the absorption of a longitudinal plasmon in the process
$\gamma_{t}+\gamma_{l}\rightarrow a$. At the symmetric position in
the upper half plane there is another pole corresponding to the emission
of a plasmon in the process $\gamma_{t}\rightarrow\gamma_{l}+a$.
As long as kinematically allowed, such collective behavior may significantly
alter the fixed-electron picture of the axion production process.
$\mathcal{F}\left(Q,\omega\right)$ provides a complete description
of the longitudinal plasmon behavior far beyond the approximated dispersion
relation $\omega^{2}\backsimeq\omega_{p}^{2}+3\left(T_{\odot}/m_{e}\right)Q^{2}$~\citep{Raffelt:1985nk}.

Based on the axion production rate of Eq.~(\ref{eq:Primakoff rate}),
the differential axion flux reaching Earth can then be written as
the convolution of the differential transition rate with the photon
blackbody distribution in the Sun,
\begin{eqnarray}
\frac{\mathrm{d}\Phi_{a}\left(E_{a}\right)}{\mathrm{d}E_{a}} & = & \frac{1}{4\pi d_{\odot}^{2}}\int_{0}^{R_{\odot}}\mathrm{d}^{3}r\int\frac{\mathrm{d}E_{\gamma}}{\pi^{2}}\frac{E_{\gamma}^{2}}{e^{E_{\gamma}/T_{\odot}}-1}\frac{\mathrm{d}\Gamma\left(\mathbf{p}_{\gamma}\right)}{\mathrm{d}E_{a}},\nonumber \\
\label{eq:total Spectrum}
\end{eqnarray}
with the Sun-Earth distance $d_{\odot}$ and the solar radius $R_{\odot}$.

In the static screening prescription, since the energy of the incident
photon equals that of the axion, the differential axion flux is given
as~\citep{Raffelt:1996wa}
\begin{eqnarray}
\frac{\mathrm{d}\Phi_{a}\left(E_{a}\right)}{\mathrm{d}E_{a}} & = & \frac{1}{4\pi d_{\odot}^{2}}\int_{0}^{R_{\odot}}\mathrm{d}^{3}r\frac{1}{\pi^{2}}\frac{E_{a}^{2}}{e^{E_{a}/T_{\odot}}-1}\Gamma_{\mathrm{s}},\label{eq:static_rate}
\end{eqnarray}
with the relevant static photon-axion conversion rate
\begin{eqnarray}
\Gamma_{\mathrm{s}} & = & \frac{T_{\odot}\kappa^{2}g_{a\gamma}^{2}}{32\pi}\left[\left(1+\frac{\kappa^{2}}{4E_{a}^{2}}\right)\ln\left(1+\frac{4E_{a}^{2}}{\kappa^{2}}\right)-1\right],\nonumber \\
\end{eqnarray}
where $\kappa^{2}=\left(4\pi\alpha/T_{\odot}\right)\left(n_{e}+\sum_{i}Z_{i}^{2}n_{N_{i}}\right)$
is the square of the Debye-Hückel scale.

\section{\label{sec:NumericalResults}Axion flux on Earth}

Equipped with the above formulation that describes the solar in-medium
effect with the linear response theory, now we are in the position
to calculate the axion flux at terrestrial detectors.

In the left panel of Fig.~\ref{fig:ElectronSpectrum} we compare
the solar Primakoff axion fluxes on Earth computed with the linear
response theory in Eq.~(\ref{eq:total Spectrum}) and with the static
screening description of the Coulomb interaction in Eq.~(\ref{eq:static_rate}).

These spectra are obtained by integrating the contributions from the
charged particles in every thin shell in the Sun, based on the Standard
Sun Model~AGSS09~\citep{Serenelli:2009yc}. In practice, the solar
radius is discretized into 100 slices, and 29 most common solar elements
are included in our computation. Besides, we assume that these solar
elements are fully ionized.
\begin{figure*}[t]
\begin{centering}
\includegraphics[scale=0.75]{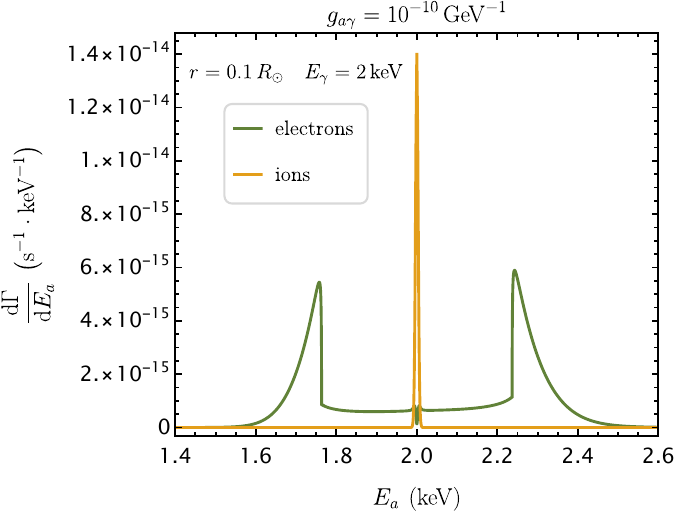}\hspace{0.8cm}\includegraphics[scale=0.74]{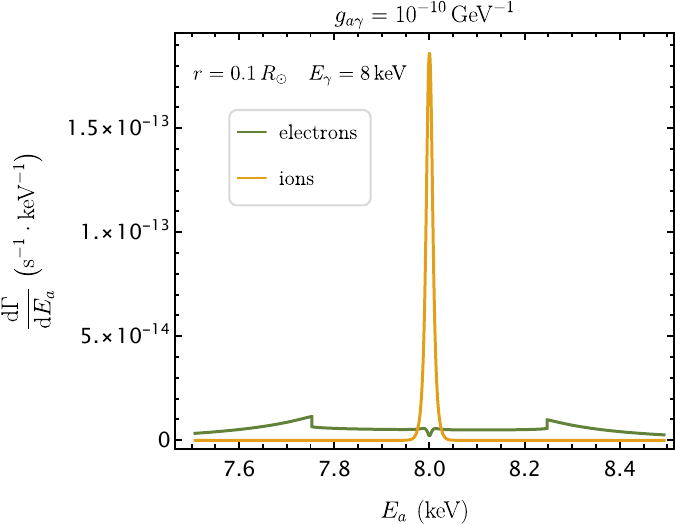}
\par\end{centering}
\caption{\label{fig:ConversionRate}The Primakoff differential photon-axion
conversion rates for photon energies $E_{\gamma}=2\,\mathrm{keV}$
(left) and $8\,\mathrm{keV}$ (right) at the solar radius $r=0.1\,R_{\odot}$,
with contributions from electrons (green) and ions (orange), respectively.
The plasmon absorption and emission peaks are clearly seen.}
\end{figure*}

While in the left panel of Fig.~\ref{fig:ElectronSpectrum} it is
observed that the average ($4.2\,\mathrm{keV}$) and the maximum ($3.0\,\mathrm{keV}$)
of the Primakoff axion energy distribution remain unchanged, the differential
rate is only $1$\textasciitilde$2\%$ lower than the calculation
based on the static structure factor in the energy range from $2$
to $12\,\mathrm{keV}$. It is quite a surprising result, given that
the denominator term $\left|1-V_{e}\,\Pi_{e}-V_{e}\,\sum_{i}{\displaystyle Z_{i}^{2}\,\Pi_{N_{i}}}\right|^{-2}$
in Eq.~(\ref{eq:Primakoff rate}) asymptotes to the Debye screening
form $\left(1+\kappa^{2}/Q^{2}\right)^{-2}$ in the limit $\omega\rightarrow0$,
where it brings a stronger screening than the static structure factor
$\left(1+\kappa^{2}/Q^{2}\right)^{-1}$ in Eq.~(\ref{eq:raffelt});
the contributions of the recoil effect and the collective effect must
coincidentally make up this loss to keep the total rate unchanged.

Such a coincidence would be difficult if there were no intrinsic relation
protecting the total rate, especially considering that in contrast
to the static screening picture, where the contributions from the
electrons and ions are in scale to the electric charge densities $n_{e}$
and $\sum_{i}Z_{i}^{2}n_{N_{i}}$ and hence a larger part of the conversion
comes from the scattering with ions, it turns out that the collective
electrons contribute dominantly to the total Primakoff axion flux
in the dynamic screening picture. Thus, our results actually confirm
 the validity of the Kramers-Kronig sum rule argument in Ref.~\citep{Raffelt:1987np},
up to a percent-level correction.

In the right panel of Fig.~\ref{fig:ElectronSpectrum} we also present
the differential solar axion flux (using the dynamic structure factor)
as an apparent surface luminosity $\phi_{a}\left(E_{a},\rho\right)$
of the solar disc~\citep{CAST:2007jps,Hoof:2021mld,Hoof:2023jol},
\begin{eqnarray}
\phi_{a}\left(E_{a},\rho\right) & = & \frac{R_{\odot}^{3}}{2\pi^{3}d_{\odot}^{2}}\int_{\rho}^{1}\frac{\tilde{r}\,\mathrm{d}\tilde{r}}{\sqrt{\tilde{r}^{2}-\rho^{2}}}\nonumber \\
 &  & \times\int\frac{\mathrm{d}E_{\gamma}}{\pi^{2}}\frac{E_{\gamma}^{2}}{e^{E_{\gamma}/T_{\odot}}-1}\frac{\mathrm{d}\Gamma\left(\mathbf{p}_{\gamma}\right)}{\mathrm{d}E_{a}},
\end{eqnarray}
where the dimensionless quantities $\tilde{r}=r/R_{\odot}$ and $\rho$
represent the radial position of the conversion process, and the distance
from the center of the solar disc, respectively.

\section{\label{sec:conclusions}discussions and conclusions}

In order to further explore the many-body effect in the solar medium
in detail, in Fig.~\ref{fig:ConversionRate} we present the differential
axion production rates for photon energies $E_{\gamma}=2\,\mathrm{keV}$
and $8\,\mathrm{keV}$ at the solar radius $r=0.1\,R_{\odot}$, respectively,
with the benchmark coupling $g_{a\gamma}=10^{-10}\mathrm{GeV}^{-1}$.
While the spectra of heavy ions are found to narrowly center at the
photon energies, behaving like static targets, it is intriguingly
observed that the non-negligible electron movement in the inner part
of the Sun can bring an energy shift up to $\mathcal{O}\left(0.1\right)\,\mathrm{keV}$
from the initial photon energies. For one thing, the two peaks in
Fig.~\ref{fig:ConversionRate} correspond to the absorption and emission
of a longitudinal plasmon at $\omega\simeq\pm\sqrt{4\pi\alpha n_{e}/m_{e}}$
in the Sun, respectively. That is, a considerable part of Primakoff
processes proceed in company with absorbing and emitting a plasmon.
For another, a broadening width of around $0.4$ keV is also clearly
seen due to the thermal movement of the electrons. While such a finite
spread of the photon energy may not bring a noticeable change to the
total spectrum of the solar axion, the strength of the resonance,
i.e., the implication of the collective effect, can only be determined
by concrete calculation.

To summarize, in this paper we have applied the linear response theory
formalism for a delicate estimate of the Primakoff photon-axion conversion
rate in the Sun. Based on this method, progress is gained in two aspects:
(1) we provide an up-to-date panoramic description of the dynamic
Primakoff process, which is explicitly shown as a combination of the
decay process $\gamma_{t}\rightarrow\gamma_{l}+a$, the plasma coalescence
process $\gamma_{t}+\gamma_{l}\rightarrow a$, and the individual
Primakoff process $\gamma_{t}+e/N\rightarrow a+e/N$; (2) beyond the
approximate Kramers-Kronig sum rules, we numerically calculate the
relevant terrestrial axion flux due to the Primakoff process, and
the flux is found to be around 1\textasciitilde 2\% lower than the
previous estimation based on the static structure factor.

Lastly, this dynamic response-oriented approach can be further applied
to other axion production mechanisms such as electron- and ion-bremsstrahlung
processes in the Sun, where a systematic treatment of the screening
and collective effects are also highly useful.\vspace{1cm}

\appendix
\begin{spacing}{1.2}

\section{\label{sec:AppendixA}Formulation for the Primakoff scattering event
rate in the Sun}
\end{spacing}

In this appendix we give a detailed derivation of the formulas in
the main text that describe the Primakoff photon-axion conversion
process in the Sun. In the nonrelativistic regime, it is convenient
to discuss in the \textit{Coulomb gauge}.

We start with the $\mathcal{T}$-matrix for the Primakoff process
where a photon (with momentum $\mathbf{p}_{\gamma}$ and polarization
$\lambda$) scatters with a nonrelativistically moving electron (illustrated
in Fig.~\ref{fig:Primakoff}), emitting an axion with momentum $\mathbf{p}_{a}$,
i.e.,
\begin{align}
 & \braket{\mathbf{p}_{a},i|\,i\mathcal{T}\,|\mathbf{p}_{\gamma},\lambda;j}=ig_{a\gamma}\,e\,\frac{\left(\mathbf{p}_{\gamma}\times\mathbf{p}_{a}\right)\cdot\hat{\boldsymbol{\varepsilon}}^{\lambda}\left(\mathbf{p}_{\gamma}\right)}{\left|\mathbf{p}_{\gamma}-\mathbf{p}_{a}\right|^{2}}\nonumber \\
 & \times\braket{i|e^{i\left(\mathbf{p}_{\gamma}-\mathbf{p}_{a}\right)\cdot\hat{\mathbf{x}}}|j}2\pi\delta\left(E_{\gamma}-E_{a}-\varepsilon_{i}+\varepsilon_{j}\right),
\end{align}
where $\hat{\boldsymbol{\varepsilon}}^{\lambda}\left(\mathbf{p}_{\gamma}\right)$
is the polarization vector for the incident photon, which satisfies
the complete and orthonormal relation, i.e., $\sum_{\lambda=\pm1}\hat{\boldsymbol{\varepsilon}}^{\lambda i}\left(\mathbf{p}_{\gamma}\right)\hat{\boldsymbol{\varepsilon}}^{\lambda j*}\left(\mathbf{p}_{\gamma}\right)=\delta^{ij}-p_{\gamma}^{i}p_{\gamma}^{j}/\left|\mathbf{p}_{\gamma}\right|^{2}$,
and $\mathbf{p}_{\gamma}\cdot\hat{\boldsymbol{\varepsilon}}^{\pm1}\left(\mathbf{p}_{\gamma}\right)=0$;
$E_{\gamma}$, $E_{a}$, $\varepsilon_{i}$, and $\varepsilon_{j}$
represent the energies of the photon, the emitted axion, the initial
and the final state of the electron, respectively.

Then we take into account the many-body effect of the solar medium
with the approach adopted in Refs.~\citep{Liang:2021zkg,Liang:2023ira}.
To this end, we resort to the linear response theory, whereby the
Primakoff event rate for a photon with momentum $\mathbf{p}_{\gamma}$
(by averaging over the initial states and summing over the final states)
is written as the following (for simplicity, here we assume only one
type of ions with charge $Ze$ and mass $m_{N}$ are present):\begin{widetext}
\begin{eqnarray}
\Gamma\left(\mathbf{p}_{\gamma}\right) & = & \sum_{i,j}\int\mathrm{d}\omega\,\delta\left(\omega-\varepsilon_{i}+\varepsilon_{j}\right)\int\mathrm{d}^{3}Q\,\delta^{(3)}\left(\mathbf{Q}+\mathbf{\mathbf{p}}_{a}-\mathbf{\mathbf{p}}_{\gamma}\right)\int\frac{\mathrm{d}^{3}p_{a}}{2E_{a}\left(2\pi\right)^{3}}\left(\frac{g_{a\gamma}\,e}{Q^{2}}\right)^{2}\frac{1}{2}\sum_{\lambda=\pm1}\frac{\left|\left(\mathbf{p}_{\gamma}\times\mathbf{p}_{a}\right)\cdot\hat{\boldsymbol{\varepsilon}}^{\lambda}\left(\mathbf{p}_{\gamma}\right)\right|^{2}}{2E_{\gamma}}\nonumber \\
 &  & \times\frac{1}{V}\int_{V}\mathrm{d}^{3}x\,\mathrm{d}^{3}x'\,p_{j}\braket{j|e^{-i\mathbf{\mathbf{\mathbf{Q}}}\cdot\mathbf{x}}\left[\hat{\rho}_{e}+\left(-Z\right)\hat{\rho}_{N}\right]\left(\mathbf{x}\right)|i}\braket{i|e^{i\mathbf{\mathbf{\mathbf{Q}}}\cdot\mathbf{x}'}\left[\hat{\rho}_{e}+\left(-Z\right)\hat{\rho}_{N}\right]\left(\mathbf{x}'\right)|j}\,2\pi\delta\left(E_{\gamma}-E_{a}-\omega\right)\nonumber \\
\nonumber \\
 & = & \int\mathrm{d}\omega\,\int\frac{\mathrm{d}^{3}Q}{\left(2\pi\right)^{3}}\frac{4\pi\alpha}{Q^{4}}\frac{g_{a\gamma}^{2}\,\left|\mathbf{p}_{\gamma}\times\mathbf{Q}\right|^{2}}{8\,E_{\gamma}\,\sqrt{\left|\mathbf{p}_{\gamma}-\mathbf{Q}\right|^{2}+m_{a}^{2}}}\frac{1}{V}\int_{V}\mathrm{d}^{3}x\,\mathrm{d}^{3}x'\,e^{-i\mathbf{\mathbf{\mathbf{Q}}}\cdot\left(\mathbf{x}-\mathbf{x}'\right)}\int_{-\infty}^{+\infty}e^{i\omega\left(0-t\right)}\mathrm{d}t\left[\braket{\hat{\rho}_{eI}\left(\mathbf{x},0\right)\hat{\rho}_{eI}\left(\mathbf{x}',t\right)}\right.\nonumber \\
 &  & \left.+\left(-Z\right)\braket{\hat{\rho}_{eI}\left(\mathbf{x},0\right)\hat{\rho}_{NI}\left(\mathbf{x}',t\right)}+\left(-Z\right)\braket{\hat{\rho}_{NI}\left(\mathbf{x},0\right)\hat{\rho}_{eI}\left(\mathbf{x}',t\right)}+\left(-Z\right)^{2}\braket{\hat{\rho}_{NI}\left(\mathbf{x},0\right)\hat{\rho}_{NI}\left(\mathbf{x}',t\right)}\right]\,\delta\left(E_{\gamma}-E_{a}-\omega\right)\nonumber \\
\nonumber \\
 & = & \int\mathrm{d}\omega\,\int\frac{\mathrm{d}^{3}Q}{\left(2\pi\right)^{3}}\frac{4\pi\alpha}{Q^{4}}\frac{g_{a\gamma}^{2}\,\left|\mathbf{p}_{\gamma}\times\mathbf{Q}\right|^{2}}{8\,E_{\gamma}\,\sqrt{\left|\mathbf{p}_{\gamma}-\mathbf{Q}\right|^{2}+m_{a}^{2}}}\frac{\left(-2\right)}{1-e^{-\omega/T_{\odot}}}\mathrm{Im}\left[\chi_{\hat{\rho}_{e}\hat{\rho}_{e}}^{\mathrm{r}}+\left(-Z\right)\chi_{\hat{\rho}_{e}\hat{\rho}_{N}}^{\mathrm{r}}+\left(-Z\right)\chi_{\hat{\rho}_{N}\hat{\rho}_{e}}^{\mathrm{r}}+\left(-Z\right)^{2}\chi_{\hat{\rho}_{N}\hat{\rho}_{N}}^{\mathrm{r}}\right]\nonumber \\
 &  & \times\delta\left(\sqrt{\left|\mathbf{p}_{\gamma}-\mathbf{Q}\right|^{2}+m_{a}^{2}}-E_{\gamma}+\omega\right)\nonumber \\
\nonumber \\
 & = & \int\mathrm{d}\omega\,\int\frac{\mathrm{d}^{3}Q}{\left(2\pi\right)^{2}}\frac{4\pi\alpha}{Q^{4}}\frac{g_{a\gamma}^{2}\,\left|\mathbf{p}_{\gamma}\times\mathbf{Q}\right|^{2}}{8\,E_{\gamma}\,\sqrt{\left|\mathbf{p}_{\gamma}-\mathbf{Q}\right|^{2}+m_{a}^{2}}}\frac{\left(-2\right)}{1-e^{-\omega/T_{\odot}}}\left[\frac{\mathrm{Im}\left(\Pi_{e}\right)}{\left|1-V_{e}\,\Pi_{e}-V_{e}\,{\displaystyle Z^{2}\,\Pi_{N}}\right|^{2}}+\frac{Z^{2}\,\mathrm{Im}\left(\Pi_{N}\right)}{\left|1-V_{e}\,\Pi_{e}-V_{e}\,{\displaystyle Z^{2}\,\Pi_{N}}\right|^{2}}\right]\nonumber \\
 &  & \times\delta\left(\sqrt{\left|\mathbf{p}_{\gamma}-\mathbf{Q}\right|^{2}+m_{a}^{2}}-E_{\gamma}+\omega\right),\label{eq:gamma0}
\end{eqnarray}
\end{widetext}where we introduce the density operators for electrons
$\hat{\rho}_{e}\left(\mathbf{x}\right)\equiv\hat{\psi}_{e}^{\dagger}\left(\mathbf{x}\right)\hat{\psi}_{e}\left(\mathbf{x}\right)$
and ions $\hat{\rho}_{N}\left(\mathbf{x}\right)\equiv\hat{\psi}_{N}^{\dagger}\left(\mathbf{x}\right)\hat{\psi}_{N}\left(\mathbf{x}\right)$,
$p_{j}$ represents the thermal distribution of the initial state
$\ket{j}$, the symbol $\left\langle \cdots\right\rangle $ represents
the thermal average, $\hat{\rho}_{eI}\left(\mathbf{x}',t\right)\equiv e^{i\hat{H}_{0}t}\hat{\rho}_{e}\left(\mathbf{x}'\right)e^{-i\hat{H}_{0}t}$
($\hat{\rho}_{NI}\left(\mathbf{x},t\right)\equiv e^{i\hat{H}_{0}t}\hat{\rho}_{N}\left(\mathbf{x}\right)e^{-i\hat{H}_{0}t}$),
with $\hat{H}_{0}$ being the unperturbed Hamiltonian of the medium
system, and $V$ is the volume of the solar medium under consideration,
which is only an intermediate quantity and is canceled in the final
expression of the event rate.

Besides, in the above derivation we invoke the fluctuation-dissipation
theorem
\begin{align}
 & S_{\hat{\rho}\hat{\rho}}\left(\mathbf{\mathbf{Q}},\,\omega\right)=\frac{1}{V}\int_{V}\mathrm{d}^{3}x\,\mathrm{d}^{3}x'\,e^{-i\mathbf{\mathbf{\mathbf{Q}}}\cdot\left(\mathbf{x}-\mathbf{x}'\right)}\int_{-\infty}^{+\infty}\mathrm{d}t\,e^{i\omega\left(0-t\right)}\nonumber \\
 & \times\braket{\hat{\rho}_{I}\left(\mathbf{x},0\right)\hat{\rho}_{I}\left(\mathbf{x}',t\right)}\nonumber \\
 & =i\frac{\left[\chi_{\hat{\rho}\hat{\rho}}\left(\mathbf{\mathbf{Q}},\,\omega+i0^{+}\right)-\chi_{\hat{\rho}\hat{\rho}}\left(\mathbf{\mathbf{Q}},\,\omega-i0^{+}\right)\right]}{1-e^{-\omega/T}}\nonumber \\
 & =\frac{-2\,\mathrm{Im}\left[\chi_{\hat{\rho}\hat{\rho}}^{\mathrm{r}}\left(\mathbf{\mathbf{Q}},\,\omega\right)\right]}{1-e^{-\omega/T}},
\end{align}
where $T$ represents temperature, $\hat{\rho}$ generally stands
for $\hat{\rho}_{e}$ and $\hat{\rho}_{N}$, so $S_{\hat{\rho}\hat{\rho}}\left(\mathbf{\mathbf{Q}},\,\omega\right)$
represents the dynamic structure factor associated with the density-density
correlation. In practice~\citep{bruus2004many}, one first evaluates
the master function $\chi_{\hat{\rho}\hat{\rho}}\left(\mathbf{Q},\,z\right)$
using the Matsubara Green's function within the framework of finite
temperature field theory, and then obtains the retarded polarizability
function $\chi_{\hat{\rho}\hat{\rho}}^{\mathrm{r}}\left(\mathbf{\mathbf{Q}},\,\omega\right)$
by performing the analytic continuation $\chi_{\hat{\rho}\hat{\rho}}^{\mathrm{r}}\left(\mathbf{\mathbf{Q}},\omega\right)=\chi_{\hat{\rho}\hat{\rho}}\left(\mathbf{\mathbf{Q}},z\rightarrow\omega+i0^{+}\right)$.

Here we take the retarded correlation function $\chi_{\hat{\rho}_{e}\hat{\rho}_{N}}^{\mathrm{r}}$
as an example to illustrate how the calculation is carried out, which
is presented as the sum of all possible diagrams that connect the
two density operators as follows, 
\tikzset{ pattern size/.store in=\mcSize,
pattern size = 5pt,
pattern thickness/.store in=\mcThickness,
pattern thickness = 0.3pt,
pattern radius/.store in=\mcRadius,
pattern radius = 1pt}
\makeatletter

\pgfutil@ifundefined{pgf@pattern@name@_fm7niqlyo}{ \pgfdeclarepatternformonly[\mcThickness,\mcSize]{_fm7niqlyo} {\pgfqpoint{0pt}{0pt}} {\pgfpoint{\mcSize}{\mcSize}} {\pgfpoint{\mcSize}{\mcSize}} { \pgfsetcolor{\tikz@pattern@color} \pgfsetlinewidth{\mcThickness} \pgfpathmoveto{\pgfqpoint{0pt}{\mcSize}} \pgfpathlineto{\pgfpoint{\mcSize+\mcThickness}{-\mcThickness}} \pgfpathmoveto{\pgfqpoint{0pt}{0pt}} \pgfpathlineto{\pgfpoint{\mcSize+\mcThickness}{\mcSize+\mcThickness}} \pgfusepath{stroke} }} \makeatother

\makeatother \tikzset{every picture/.style={line width=0.75pt}} 

\begin{align}
\label{mixing}
&
\chi_{\hat{\rho}_{e}\hat{\rho}_{N}}^{\mathrm{r}}=\vcenter{\hbox{
\begin{tikzpicture}[x=0.20pt,y=0.20pt,yscale=-1,xscale=1]
\draw  [pattern=_fm7niqlyo,pattern size=5pt,pattern thickness=0.75pt,pattern radius=0pt, pattern color={rgb, 255:red, 0; green, 0; blue, 0}] (231.03,137) -- (400.2,137) -- (400.2,257.78) -- (231.03,257.78) -- cycle ; 
\draw   (231.03,137) -- (400.2,137) .. controls (422.18,137) and (440,164.04) .. (440,197.39) .. controls (440,230.74) and (422.18,257.78) .. (400.2,257.78) -- (231.03,257.78) .. controls (209.05,257.78) and (191.23,230.74) .. (191.23,197.39) .. controls (191.23,164.04) and (209.05,137) .. (231.03,137) -- cycle ;
\draw  [fill={rgb, 255:red, 128; green, 128; blue, 128 }  ,fill opacity=1 ] (185.73,196.39) .. controls (185.73,199.7) and (188.64,202.39) .. (192.23,202.39) .. controls (195.82,202.39) and (198.73,199.7) .. (198.73,196.39) .. controls (198.73,193.07) and (195.82,190.39) .. (192.23,190.39) .. controls (188.64,190.39) and (185.73,193.07) .. (185.73,196.39) -- cycle ;
\draw    (400,146) .. controls (440,142) and (450,240) .. (401,253) ;
\draw  [fill={rgb, 255:red, 128; green, 128; blue, 128 }  ,fill opacity=1 ] (430.73,195.39) .. controls (430.73,198.7) and (433.64,201.39) .. (437.23,201.39) .. controls (440.82,201.39) and (443.73,198.7) .. (443.73,195.39) .. controls (443.73,192.07) and (440.82,189.39) .. (437.23,189.39) .. controls (433.64,189.39) and (430.73,192.07) .. (430.73,195.39) -- cycle ;
\end{tikzpicture}
}} \nonumber    \\
\nonumber    \\
& =\vcenter{\hbox{
\begin{tikzpicture}[x=0.25pt,y=0.25pt,yscale=-1,xscale=1]
\draw   (124.36,160.23) .. controls (124.36,135.17) and (145.11,114.85) .. (170.7,114.85) .. controls (196.29,114.85) and (217.04,135.17) .. (217.04,160.23) .. controls (217.04,185.3) and (196.29,205.62) .. (170.7,205.62) .. controls (145.11,205.62) and (124.36,185.3) .. (124.36,160.23) -- cycle ;
\draw  [draw opacity=0] (184.74,204.96) .. controls (180.9,205.94) and (176.86,206.47) .. (172.7,206.47) .. controls (146.6,206.47) and (125.44,185.77) .. (125.44,160.23) .. controls (125.44,135.78) and (144.84,115.77) .. (169.4,114.11) -- (172.7,160.23) -- cycle ;
\draw    (181.77,205.62) .. controls (178.83,206.18) and (175.8,206.47) .. (172.7,206.47) .. controls (146.6,206.47) and (125.44,185.77) .. (125.44,160.23) .. controls (125.44,135.44) and (145.39,115.21) .. (170.44,114.05) ;
\draw [shift={(167.57,114.27)}, rotate = 356.14] [fill={rgb, 255:red, 0; green, 0; blue, 0 }  ][line width=0.08]  [draw opacity=0] (10.72,-5.15) -- (0,0) -- (10.72,5.15) -- (7.12,0) -- cycle    ;
\draw [shift={(184.74,204.96)}, rotate = 165.59] [fill={rgb, 255:red, 0; green, 0; blue, 0 }  ][line width=0.08]  [draw opacity=0] (10.72,-5.15) -- (0,0) -- (10.72,5.15) -- (7.12,0) -- cycle    ;
\draw  [fill={rgb, 255:red, 128; green, 128; blue, 128 }  ,fill opacity=1 ] (385,156.75) .. controls (385,159.37) and (387.24,161.5) .. (390,161.5) .. controls (392.76,161.5) and (395,159.37) .. (395,156.75) .. controls (395,154.13) and (392.76,152) .. (390,152) .. controls (387.24,152) and (385,154.13) .. (385,156.75) -- cycle ;
\draw  [line width=0.75]  (216.96,155.4) .. controls (218.85,157.65) and (220.66,159.8) .. (222.76,159.8) .. controls (224.86,159.8) and (226.67,157.65) .. (228.57,155.4) .. controls (230.46,153.14) and (232.27,151) .. (234.37,151) .. controls (236.47,151) and (238.28,153.14) .. (240.17,155.4) .. controls (242.07,157.65) and (243.88,159.8) .. (245.98,159.8) .. controls (248.08,159.8) and (249.89,157.65) .. (251.78,155.4) .. controls (253.67,153.14) and (255.48,151) .. (257.58,151) .. controls (259.69,151) and (261.5,153.14) .. (263.39,155.4) .. controls (265.28,157.65) and (267.09,159.8) .. (269.19,159.8) .. controls (271.29,159.8) and (273.1,157.65) .. (275,155.4) .. controls (276.89,153.14) and (278.7,151) .. (280.8,151) .. controls (282.9,151) and (284.71,153.14) .. (286.6,155.4) .. controls (288.5,157.65) and (290.31,159.8) .. (292.41,159.8) .. controls (294.51,159.8) and (296.32,157.65) .. (298.21,155.4) ;
\draw  [line width=0.65]  (217.75,158.79) .. controls (219.64,161.04) and (221.45,163.18) .. (223.55,163.18) .. controls (225.65,163.18) and (227.46,161.04) .. (229.35,158.79) .. controls (231.25,156.53) and (233.06,154.39) .. (235.16,154.39) .. controls (237.26,154.39) and (239.07,156.53) .. (240.96,158.79) .. controls (242.86,161.04) and (244.67,163.18) .. (246.77,163.18) .. controls (248.87,163.18) and (250.68,161.04) .. (252.57,158.79) .. controls (254.46,156.53) and (256.27,154.39) .. (258.37,154.39) .. controls (260.47,154.39) and (262.28,156.53) .. (264.18,158.79) .. controls (266.07,161.04) and (267.88,163.18) .. (269.98,163.18) .. controls (272.08,163.18) and (273.89,161.04) .. (275.78,158.79) .. controls (277.68,156.53) and (279.49,154.39) .. (281.59,154.39) .. controls (283.69,154.39) and (285.5,156.53) .. (287.39,158.79) .. controls (289.29,161.04) and (291.1,163.18) .. (293.2,163.18) .. controls (295.3,163.18) and (297.11,161.04) .. (299,158.79) ;
\draw   (306.68,158.06) .. controls (306.68,136.86) and (324.06,119.68) .. (345.5,119.68) .. controls (366.94,119.68) and (384.32,136.86) .. (384.32,158.06) .. controls (384.32,179.26) and (366.94,196.45) .. (345.5,196.45) .. controls (324.06,196.45) and (306.68,179.26) .. (306.68,158.06)(299,158.06) .. controls (299,132.62) and (319.82,112) .. (345.5,112) .. controls (371.18,112) and (392,132.62) .. (392,158.06) .. controls (392,183.5) and (371.18,204.13) .. (345.5,204.13) .. controls (319.82,204.13) and (299,183.5) .. (299,158.06) ;
\draw  [fill={rgb, 255:red, 0; green, 0; blue, 0 }  ,fill opacity=1 ] (344.8,192) -- (356,200) -- (344.8,208) -- (347.71,200) -- cycle ;
\draw  [fill={rgb, 255:red, 0; green, 0; blue, 0 }  ,fill opacity=1 ] (354,123.13) -- (341,116.06) -- (354,109) -- (350.49,116.06) -- cycle ;
\draw  [fill={rgb, 255:red, 128; green, 128; blue, 128 }  ,fill opacity=1 ] (212,158.75) .. controls (212,161.37) and (214.24,163.5) .. (217,163.5) .. controls (219.76,163.5) and (222,161.37) .. (222,158.75) .. controls (222,156.13) and (219.76,154) .. (217,154) .. controls (214.24,154) and (212,156.13) .. (212,158.75) -- cycle ;
\draw  [fill={rgb, 255:red, 128; green, 128; blue, 128 }  ,fill opacity=1 ] (119,161.75) .. controls (119,164.37) and (121.24,166.5) .. (124,166.5) .. controls (126.76,166.5) and (129,164.37) .. (129,161.75) .. controls (129,159.13) and (126.76,157) .. (124,157) .. controls (121.24,157) and (119,159.13) .. (119,161.75) -- cycle ;
\draw  [fill={rgb, 255:red, 128; green, 128; blue, 128 }  ,fill opacity=1 ] (298,156.75) .. controls (298,159.37) and (300.24,161.5) .. (303,161.5) .. controls (305.76,161.5) and (308,159.37) .. (308,156.75) .. controls (308,154.13) and (305.76,152) .. (303,152) .. controls (300.24,152) and (298,154.13) .. (298,156.75) -- cycle ;
\end{tikzpicture}}}+\vcenter{\hbox{\begin{tikzpicture}[x=0.25pt,y=0.25pt,yscale=-1,xscale=1]
\draw   (291.36,97.23) .. controls (291.36,72.17) and (312.11,51.85) .. (337.7,51.85) .. controls (363.29,51.85) and (384.04,72.17) .. (384.04,97.23) .. controls (384.04,122.3) and (363.29,142.62) .. (337.7,142.62) .. controls (312.11,142.62) and (291.36,122.3) .. (291.36,97.23) -- cycle ;
\draw  [fill={rgb, 255:red, 128; green, 128; blue, 128 }  ,fill opacity=1 ] (552,93.75) .. controls (552,96.37) and (554.24,98.5) .. (557,98.5) .. controls (559.76,98.5) and (562,96.37) .. (562,93.75) .. controls (562,91.13) and (559.76,89) .. (557,89) .. controls (554.24,89) and (552,91.13) .. (552,93.75) -- cycle ;
\draw  [line width=0.75]  (383.96,92.4) .. controls (385.85,94.65) and (387.66,96.8) .. (389.76,96.8) .. controls (391.86,96.8) and (393.67,94.65) .. (395.57,92.4) .. controls (397.46,90.14) and (399.27,88) .. (401.37,88) .. controls (403.47,88) and (405.28,90.14) .. (407.17,92.4) .. controls (409.07,94.65) and (410.88,96.8) .. (412.98,96.8) .. controls (415.08,96.8) and (416.89,94.65) .. (418.78,92.4) .. controls (420.67,90.14) and (422.48,88) .. (424.58,88) .. controls (426.69,88) and (428.5,90.14) .. (430.39,92.4) .. controls (432.28,94.65) and (434.09,96.8) .. (436.19,96.8) .. controls (438.29,96.8) and (440.1,94.65) .. (442,92.4) .. controls (443.89,90.14) and (445.7,88) .. (447.8,88) .. controls (449.9,88) and (451.71,90.14) .. (453.6,92.4) .. controls (455.5,94.65) and (457.31,96.8) .. (459.41,96.8) .. controls (461.51,96.8) and (463.32,94.65) .. (465.21,92.4) ; 
\draw  [line width=0.75]  (384.75,95.79) .. controls (386.64,98.04) and (388.45,100.18) .. (390.55,100.18) .. controls (392.65,100.18) and (394.46,98.04) .. (396.35,95.79) .. controls (398.25,93.53) and (400.06,91.39) .. (402.16,91.39) .. controls (404.26,91.39) and (406.07,93.53) .. (407.96,95.79) .. controls (409.86,98.04) and (411.67,100.18) .. (413.77,100.18) .. controls (415.87,100.18) and (417.68,98.04) .. (419.57,95.79) .. controls (421.46,93.53) and (423.27,91.39) .. (425.37,91.39) .. controls (427.47,91.39) and (429.28,93.53) .. (431.18,95.79) .. controls (433.07,98.04) and (434.88,100.18) .. (436.98,100.18) .. controls (439.08,100.18) and (440.89,98.04) .. (442.78,95.79) .. controls (444.68,93.53) and (446.49,91.39) .. (448.59,91.39) .. controls (450.69,91.39) and (452.5,93.53) .. (454.39,95.79) .. controls (456.29,98.04) and (458.1,100.18) .. (460.2,100.18) .. controls (462.3,100.18) and (464.11,98.04) .. (466,95.79) ;
\draw   (473.68,95.06) .. controls (473.68,73.86) and (491.06,56.68) .. (512.5,56.68) .. controls (533.94,56.68) and (551.32,73.86) .. (551.32,95.06) .. controls (551.32,116.26) and (533.94,133.45) .. (512.5,133.45) .. controls (491.06,133.45) and (473.68,116.26) .. (473.68,95.06)(466,95.06) .. controls (466,69.62) and (486.82,49) .. (512.5,49) .. controls (538.18,49) and (559,69.62) .. (559,95.06) .. controls (559,120.5) and (538.18,141.13) .. (512.5,141.13) .. controls (486.82,141.13) and (466,120.5) .. (466,95.06) ; \draw  [fill={rgb, 255:red, 0; green, 0; blue, 0 }  ,fill opacity=1 ] (511.8,129) -- (523,137) -- (511.8,145) -- (514.71,137) -- cycle ; \draw  [fill={rgb, 255:red, 0; green, 0; blue, 0 }  ,fill opacity=1 ] (521,60.13) -- (508,53.06) -- (521,46) -- (517.49,53.06) -- cycle ;
\draw  [fill={rgb, 255:red, 128; green, 128; blue, 128 }  ,fill opacity=1 ] (379,95.75) .. controls (379,98.37) and (381.24,100.5) .. (384,100.5) .. controls (386.76,100.5) and (389,98.37) .. (389,95.75) .. controls (389,93.13) and (386.76,91) .. (384,91) .. controls (381.24,91) and (379,93.13) .. (379,95.75) -- cycle ;
\draw  [fill={rgb, 255:red, 128; green, 128; blue, 128 }  ,fill opacity=1 ] (286,98.75) .. controls (286,101.37) and (288.24,103.5) .. (291,103.5) .. controls (293.76,103.5) and (296,101.37) .. (296,98.75) .. controls (296,96.13) and (293.76,94) .. (291,94) .. controls (288.24,94) and (286,96.13) .. (286,98.75) -- cycle ;
\draw  [fill={rgb, 255:red, 128; green, 128; blue, 128 }  ,fill opacity=1 ] (465,93.75) .. controls (465,96.37) and (467.24,98.5) .. (470,98.5) .. controls (472.76,98.5) and (475,96.37) .. (475,93.75) .. controls (475,91.13) and (472.76,89) .. (470,89) .. controls (467.24,89) and (465,91.13) .. (465,93.75) -- cycle ;
\draw   (114.36,98.23) .. controls (114.36,73.17) and (135.11,52.85) .. (160.7,52.85) .. controls (186.29,52.85) and (207.04,73.17) .. (207.04,98.23) .. controls (207.04,123.3) and (186.29,143.62) .. (160.7,143.62) .. controls (135.11,143.62) and (114.36,123.3) .. (114.36,98.23) -- cycle ;
\draw  [draw opacity=0] (174.74,142.96) .. controls (170.9,143.94) and (166.86,144.47) .. (162.7,144.47) .. controls (136.6,144.47) and (115.44,123.77) .. (115.44,98.23) .. controls (115.44,73.78) and (134.84,53.77) .. (159.4,52.11) -- (162.7,98.23) -- cycle ; \draw    (171.77,143.62) .. controls (168.83,144.18) and (165.8,144.47) .. (162.7,144.47) .. controls (136.6,144.47) and (115.44,123.77) .. (115.44,98.23) .. controls (115.44,73.44) and (135.39,53.21) .. (160.44,52.05) ; \draw [shift={(157.57,52.27)}, rotate = 356.14] [fill={rgb, 255:red, 0; green, 0; blue, 0 }  ][line width=0.08]  [draw opacity=0] (10.72,-5.15) -- (0,0) -- (10.72,5.15) -- (7.12,0) -- cycle    ; \draw [shift={(174.74,142.96)}, rotate = 165.59] [fill={rgb, 255:red, 0; green, 0; blue, 0 }  ][line width=0.08]  [draw opacity=0] (10.72,-5.15) -- (0,0) -- (10.72,5.15) -- (7.12,0) -- cycle    ;
\draw  [line width=0.75]  (206.96,93.4) .. controls (208.85,95.65) and (210.66,97.8) .. (212.76,97.8) .. controls (214.86,97.8) and (216.67,95.65) .. (218.57,93.4) .. controls (220.46,91.14) and (222.27,89) .. (224.37,89) .. controls (226.47,89) and (228.28,91.14) .. (230.17,93.4) .. controls (232.07,95.65) and (233.88,97.8) .. (235.98,97.8) .. controls (238.08,97.8) and (239.89,95.65) .. (241.78,93.4) .. controls (243.67,91.14) and (245.48,89) .. (247.58,89) .. controls (249.69,89) and (251.5,91.14) .. (253.39,93.4) .. controls (255.28,95.65) and (257.09,97.8) .. (259.19,97.8) .. controls (261.29,97.8) and (263.1,95.65) .. (265,93.4) .. controls (266.89,91.14) and (268.7,89) .. (270.8,89) .. controls (272.9,89) and (274.71,91.14) .. (276.6,93.4) .. controls (278.5,95.65) and (280.31,97.8) .. (282.41,97.8) .. controls (284.51,97.8) and (286.32,95.65) .. (288.21,93.4) ; 
\draw  [line width=0.75]  (207.75,96.79) .. controls (209.64,99.04) and (211.45,101.18) .. (213.55,101.18) .. controls (215.65,101.18) and (217.46,99.04) .. (219.35,96.79) .. controls (221.25,94.53) and (223.06,92.39) .. (225.16,92.39) .. controls (227.26,92.39) and (229.07,94.53) .. (230.96,96.79) .. controls (232.86,99.04) and (234.67,101.18) .. (236.77,101.18) .. controls (238.87,101.18) and (240.68,99.04) .. (242.57,96.79) .. controls (244.46,94.53) and (246.27,92.39) .. (248.37,92.39) .. controls (250.47,92.39) and (252.28,94.53) .. (254.18,96.79) .. controls (256.07,99.04) and (257.88,101.18) .. (259.98,101.18) .. controls (262.08,101.18) and (263.89,99.04) .. (265.78,96.79) .. controls (267.68,94.53) and (269.49,92.39) .. (271.59,92.39) .. controls (273.69,92.39) and (275.5,94.53) .. (277.39,96.79) .. controls (279.29,99.04) and (281.1,101.18) .. (283.2,101.18) .. controls (285.3,101.18) and (287.11,99.04) .. (289,96.79) ;
\draw  [fill={rgb, 255:red, 128; green, 128; blue, 128 }  ,fill opacity=1 ] (202,96.75) .. controls (202,99.37) and (204.24,101.5) .. (207,101.5) .. controls (209.76,101.5) and (212,99.37) .. (212,96.75) .. controls (212,94.13) and (209.76,92) .. (207,92) .. controls (204.24,92) and (202,94.13) .. (202,96.75) -- cycle ;
\draw  [fill={rgb, 255:red, 128; green, 128; blue, 128 }  ,fill opacity=1 ] (109,99.75) .. controls (109,102.37) and (111.24,104.5) .. (114,104.5) .. controls (116.76,104.5) and (119,102.37) .. (119,99.75) .. controls (119,97.13) and (116.76,95) .. (114,95) .. controls (111.24,95) and (109,97.13) .. (109,99.75) -- cycle ;
\draw  [draw opacity=0] (351.74,141.96) .. controls (347.9,142.94) and (343.86,143.47) .. (339.7,143.47) .. controls (313.6,143.47) and (292.44,122.77) .. (292.44,97.23) .. controls (292.44,72.78) and (311.84,52.77) .. (336.4,51.11) -- (339.7,97.23) -- cycle ; \draw    (348.77,142.62) .. controls (345.83,143.18) and (342.8,143.47) .. (339.7,143.47) .. controls (313.6,143.47) and (292.44,122.77) .. (292.44,97.23) .. controls (292.44,72.44) and (312.39,52.21) .. (337.44,51.05) ; \draw [shift={(334.57,51.27)}, rotate = 356.14] [fill={rgb, 255:red, 0; green, 0; blue, 0 }  ][line width=0.08]  [draw opacity=0] (10.72,-5.15) -- (0,0) -- (10.72,5.15) -- (7.12,0) -- cycle    ; \draw [shift={(351.74,141.96)}, rotate = 165.59] [fill={rgb, 255:red, 0; green, 0; blue, 0 }  ][line width=0.08]  [draw opacity=0] (10.72,-5.15) -- (0,0) -- (10.72,5.15) -- (7.12,0) -- cycle    ;
\end{tikzpicture} }}+\cdots\nonumber \\
\nonumber \\
&=\frac{\vcenter{\hbox{\begin{tikzpicture}[x=0.25pt,y=0.25pt,yscale=-1,xscale=1]
\draw   (124.36,160.23) .. controls (124.36,135.17) and (145.11,114.85) .. (170.7,114.85) .. controls (196.29,114.85) and (217.04,135.17) .. (217.04,160.23) .. controls (217.04,185.3) and (196.29,205.62) .. (170.7,205.62) .. controls (145.11,205.62) and (124.36,185.3) .. (124.36,160.23) -- cycle ;
\draw  [draw opacity=0] (184.74,204.96) .. controls (180.9,205.94) and (176.86,206.47) .. (172.7,206.47) .. controls (146.6,206.47) and (125.44,185.77) .. (125.44,160.23) .. controls (125.44,135.78) and (144.84,115.77) .. (169.4,114.11) -- (172.7,160.23) -- cycle ;
\draw    (181.77,205.62) .. controls (178.83,206.18) and (175.8,206.47) .. (172.7,206.47) .. controls (146.6,206.47) and (125.44,185.77) .. (125.44,160.23) .. controls (125.44,135.44) and (145.39,115.21) .. (170.44,114.05) ;
\draw [shift={(167.57,114.27)}, rotate = 356.14] [fill={rgb, 255:red, 0; green, 0; blue, 0 }  ][line width=0.08]  [draw opacity=0] (10.72,-5.15) -- (0,0) -- (10.72,5.15) -- (7.12,0) -- cycle    ;
\draw [shift={(184.74,204.96)}, rotate = 165.59] [fill={rgb, 255:red, 0; green, 0; blue, 0 }  ][line width=0.08]  [draw opacity=0] (10.72,-5.15) -- (0,0) -- (10.72,5.15) -- (7.12,0) -- cycle    ;
\draw  [fill={rgb, 255:red, 128; green, 128; blue, 128 }  ,fill opacity=1 ] (385,156.75) .. controls (385,159.37) and (387.24,161.5) .. (390,161.5) .. controls (392.76,161.5) and (395,159.37) .. (395,156.75) .. controls (395,154.13) and (392.76,152) .. (390,152) .. controls (387.24,152) and (385,154.13) .. (385,156.75) -- cycle ;
\draw  [line width=0.75]  (216.96,155.4) .. controls (218.85,157.65) and (220.66,159.8) .. (222.76,159.8) .. controls (224.86,159.8) and (226.67,157.65) .. (228.57,155.4) .. controls (230.46,153.14) and (232.27,151) .. (234.37,151) .. controls (236.47,151) and (238.28,153.14) .. (240.17,155.4) .. controls (242.07,157.65) and (243.88,159.8) .. (245.98,159.8) .. controls (248.08,159.8) and (249.89,157.65) .. (251.78,155.4) .. controls (253.67,153.14) and (255.48,151) .. (257.58,151) .. controls (259.69,151) and (261.5,153.14) .. (263.39,155.4) .. controls (265.28,157.65) and (267.09,159.8) .. (269.19,159.8) .. controls (271.29,159.8) and (273.1,157.65) .. (275,155.4) .. controls (276.89,153.14) and (278.7,151) .. (280.8,151) .. controls (282.9,151) and (284.71,153.14) .. (286.6,155.4) .. controls (288.5,157.65) and (290.31,159.8) .. (292.41,159.8) .. controls (294.51,159.8) and (296.32,157.65) .. (298.21,155.4) ;
\draw  [line width=0.65]  (217.75,158.79) .. controls (219.64,161.04) and (221.45,163.18) .. (223.55,163.18) .. controls (225.65,163.18) and (227.46,161.04) .. (229.35,158.79) .. controls (231.25,156.53) and (233.06,154.39) .. (235.16,154.39) .. controls (237.26,154.39) and (239.07,156.53) .. (240.96,158.79) .. controls (242.86,161.04) and (244.67,163.18) .. (246.77,163.18) .. controls (248.87,163.18) and (250.68,161.04) .. (252.57,158.79) .. controls (254.46,156.53) and (256.27,154.39) .. (258.37,154.39) .. controls (260.47,154.39) and (262.28,156.53) .. (264.18,158.79) .. controls (266.07,161.04) and (267.88,163.18) .. (269.98,163.18) .. controls (272.08,163.18) and (273.89,161.04) .. (275.78,158.79) .. controls (277.68,156.53) and (279.49,154.39) .. (281.59,154.39) .. controls (283.69,154.39) and (285.5,156.53) .. (287.39,158.79) .. controls (289.29,161.04) and (291.1,163.18) .. (293.2,163.18) .. controls (295.3,163.18) and (297.11,161.04) .. (299,158.79) ;
\draw   (306.68,158.06) .. controls (306.68,136.86) and (324.06,119.68) .. (345.5,119.68) .. controls (366.94,119.68) and (384.32,136.86) .. (384.32,158.06) .. controls (384.32,179.26) and (366.94,196.45) .. (345.5,196.45) .. controls (324.06,196.45) and (306.68,179.26) .. (306.68,158.06)(299,158.06) .. controls (299,132.62) and (319.82,112) .. (345.5,112) .. controls (371.18,112) and (392,132.62) .. (392,158.06) .. controls (392,183.5) and (371.18,204.13) .. (345.5,204.13) .. controls (319.82,204.13) and (299,183.5) .. (299,158.06) ;
\draw  [fill={rgb, 255:red, 0; green, 0; blue, 0 }  ,fill opacity=1 ] (344.8,192) -- (356,200) -- (344.8,208) -- (347.71,200) -- cycle ;
\draw  [fill={rgb, 255:red, 0; green, 0; blue, 0 }  ,fill opacity=1 ] (354,123.13) -- (341,116.06) -- (354,109) -- (350.49,116.06) -- cycle ;
\draw  [fill={rgb, 255:red, 128; green, 128; blue, 128 }  ,fill opacity=1 ] (212,158.75) .. controls (212,161.37) and (214.24,163.5) .. (217,163.5) .. controls (219.76,163.5) and (222,161.37) .. (222,158.75) .. controls (222,156.13) and (219.76,154) .. (217,154) .. controls (214.24,154) and (212,156.13) .. (212,158.75) -- cycle ;
\draw  [fill={rgb, 255:red, 128; green, 128; blue, 128 }  ,fill opacity=1 ] (119,161.75) .. controls (119,164.37) and (121.24,166.5) .. (124,166.5) .. controls (126.76,166.5) and (129,164.37) .. (129,161.75) .. controls (129,159.13) and (126.76,157) .. (124,157) .. controls (121.24,157) and (119,159.13) .. (119,161.75) -- cycle ;
\draw  [fill={rgb, 255:red, 128; green, 128; blue, 128 }  ,fill opacity=1 ] (298,156.75) .. controls (298,159.37) and (300.24,161.5) .. (303,161.5) .. controls (305.76,161.5) and (308,159.37) .. (308,156.75) .. controls (308,154.13) and (305.76,152) .. (303,152) .. controls (300.24,152) and (298,154.13) .. (298,156.75) -- cycle ;
\end{tikzpicture}}}}{1-\vcenter{\hbox{\begin{tikzpicture}[x=0.25pt,y=0.25pt,yscale=-1,xscale=1]
\draw   (312.36,757.23) .. controls (312.36,732.17) and (333.11,711.85) .. (358.7,711.85) .. controls (384.29,711.85) and (405.04,732.17) .. (405.04,757.23) .. controls (405.04,782.3) and (384.29,802.62) .. (358.7,802.62) .. controls (333.11,802.62) and (312.36,782.3) .. (312.36,757.23) -- cycle ;
\draw  [line width=0.75]  (404.96,752.4) .. controls (406.85,754.65) and (408.66,756.8) .. (410.76,756.8) .. controls (412.86,756.8) and (414.67,754.65) .. (416.57,752.4) .. controls (418.46,750.14) and (420.27,748) .. (422.37,748) .. controls (424.47,748) and (426.28,750.14) .. (428.17,752.4) .. controls (430.07,754.65) and (431.88,756.8) .. (433.98,756.8) .. controls (436.08,756.8) and (437.89,754.65) .. (439.78,752.4) .. controls (441.67,750.14) and (443.48,748) .. (445.58,748) .. controls (447.69,748) and (449.5,750.14) .. (451.39,752.4) .. controls (453.28,754.65) and (455.09,756.8) .. (457.19,756.8) .. controls (459.29,756.8) and (461.1,754.65) .. (463,752.4) .. controls (464.89,750.14) and (466.7,748) .. (468.8,748) .. controls (470.9,748) and (472.71,750.14) .. (474.6,752.4) .. controls (476.5,754.65) and (478.31,756.8) .. (480.41,756.8) .. controls (482.51,756.8) and (484.32,754.65) .. (486.21,752.4) ;
\draw  [line width=0.75]  (405.75,755.79) .. controls (407.64,758.04) and (409.45,760.18) .. (411.55,760.18) .. controls (413.65,760.18) and (415.46,758.04) .. (417.35,755.79) .. controls (419.25,753.53) and (421.06,751.39) .. (423.16,751.39) .. controls (425.26,751.39) and (427.07,753.53) .. (428.96,755.79) .. controls (430.86,758.04) and (432.67,760.18) .. (434.77,760.18) .. controls (436.87,760.18) and (438.68,758.04) .. (440.57,755.79) .. controls (442.46,753.53) and (444.27,751.39) .. (446.37,751.39) .. controls (448.47,751.39) and (450.28,753.53) .. (452.18,755.79) .. controls (454.07,758.04) and (455.88,760.18) .. (457.98,760.18) .. controls (460.08,760.18) and (461.89,758.04) .. (463.78,755.79) .. controls (465.68,753.53) and (467.49,751.39) .. (469.59,751.39) .. controls (471.69,751.39) and (473.5,753.53) .. (475.39,755.79) .. controls (477.29,758.04) and (479.1,760.18) .. (481.2,760.18) .. controls (483.3,760.18) and (485.11,758.04) .. (487,755.79) ;
\draw  [fill={rgb, 255:red, 128; green, 128; blue, 128 }  ,fill opacity=1 ] (400,755.75) .. controls (400,758.37) and (402.24,760.5) .. (405,760.5) .. controls (407.76,760.5) and (410,758.37) .. (410,755.75) .. controls (410,753.13) and (407.76,751) .. (405,751) .. controls (402.24,751) and (400,753.13) .. (400,755.75) -- cycle ; 
\draw  [fill={rgb, 255:red, 128; green, 128; blue, 128 }  ,fill opacity=1 ] (307,758.75) .. controls (307,761.37) and (309.24,763.5) .. (312,763.5) .. controls (314.76,763.5) and (317,761.37) .. (317,758.75) .. controls (317,756.13) and (314.76,754) .. (312,754) .. controls (309.24,754) and (307,756.13) .. (307,758.75) -- cycle ;
\draw  [draw opacity=0] (372.74,801.96) .. controls (368.9,802.94) and (364.86,803.47) .. (360.7,803.47) .. controls (334.6,803.47) and (313.44,782.77) .. (313.44,757.23) .. controls (313.44,732.78) and (332.84,712.77) .. (357.4,711.11) -- (360.7,757.23) -- cycle ; \draw    (369.77,802.62) .. controls (366.83,803.18) and (363.8,803.47) .. (360.7,803.47) .. controls (334.6,803.47) and (313.44,782.77) .. (313.44,757.23) .. controls (313.44,732.44) and (333.39,712.21) .. (358.44,711.05) ; \draw [shift={(355.57,711.27)}, rotate = 356.14] [fill={rgb, 255:red, 0; green, 0; blue, 0 }  ][line width=0.08]  [draw opacity=0] (10.72,-5.15) -- (0,0) -- (10.72,5.15) -- (7.12,0) -- cycle    ; \draw [shift={(372.74,801.96)}, rotate = 165.59] [fill={rgb, 255:red, 0; green, 0; blue, 0 }  ][line width=0.08]  [draw opacity=0] (10.72,-5.15) -- (0,0) -- (10.72,5.15) -- (7.12,0) -- cycle    ;
\end{tikzpicture}}}},
\end{align}where ~$\vcenter{\hbox{
\begin{tikzpicture}[x=0.27pt,y=0.27pt,yscale=-1,xscale=1]
\draw  [draw opacity=0] (311.61,107.27) .. controls (309.48,107.75) and (307.27,108) .. (305,108) .. controls (288.43,108) and (275,94.57) .. (275,78) .. controls (275,62.47) and (286.79,49.7) .. (301.91,48.16) -- (305,78) -- cycle ; \draw    (308.62,107.78) .. controls (307.43,107.93) and (306.23,108) .. (305,108) .. controls (288.43,108) and (275,94.57) .. (275,78) .. controls (275,62.13) and (287.33,49.13) .. (302.93,48.07) ; \draw [shift={(300.11,48.4)}, rotate = 354.16] [fill={rgb, 255:red, 0; green, 0; blue, 0 }  ][line width=0.08]  [draw opacity=0] (10.72,-5.15) -- (0,0) -- (10.72,5.15) -- (7.12,0) -- cycle    ; \draw [shift={(311.61,107.27)}, rotate = 167.33] [fill={rgb, 255:red, 0; green, 0; blue, 0 }  ][line width=0.08]  [draw opacity=0] (10.72,-5.15) -- (0,0) -- (10.72,5.15) -- (7.12,0) -- cycle    ;
\draw  [draw opacity=0] (306.87,48.06) .. controls (322.57,49.02) and (335,62.06) .. (335,78) .. controls (335,94.26) and (322.06,107.5) .. (305.91,107.99) -- (305,78) -- cycle ; \draw   (306.87,48.06) .. controls (322.57,49.02) and (335,62.06) .. (335,78) .. controls (335,94.26) and (322.06,107.5) .. (305.91,107.99) ;
\draw  [fill={rgb, 255:red, 128; green, 128; blue, 128 }  ,fill opacity=1 ] (272.4,79) .. controls (272.4,80.1) and (273.12,81) .. (274,81) .. controls (274.88,81) and (275.6,80.1) .. (275.6,79) .. controls (275.6,77.9) and (274.88,77) .. (274,77) .. controls (273.12,77) and (272.4,77.9) .. (272.4,79) -- cycle ;
\draw  [fill={rgb, 255:red, 128; green, 128; blue, 128 }  ,fill opacity=1 ] (335.4,79) .. controls (335.4,80.1) and (336.12,81) .. (337,81) .. controls (337.88,81) and (338.6,80.1) .. (338.6,79) .. controls (338.6,77.9) and (337.88,77) .. (337,77) .. controls (336.12,77) and (335.4,77.9) .. (335.4,79) -- cycle ;
\end{tikzpicture}}}$ and ~$\vcenter{\hbox{
\begin{tikzpicture}[x=0.18pt,y=0.18pt,yscale=-1,xscale=1]
\draw  [fill={rgb, 255:red, 128; green, 128; blue, 128 }  ,fill opacity=1 ] (351,123.75) .. controls (351,126.37) and (353.24,128.5) .. (356,128.5) .. controls (358.76,128.5) and (361,126.37) .. (361,123.75) .. controls (361,121.13) and (358.76,119) .. (356,119) .. controls (353.24,119) and (351,121.13) .. (351,123.75) -- cycle ; 
\draw   (272.68,125.06) .. controls (272.68,103.86) and (290.06,86.68) .. (311.5,86.68) .. controls (332.94,86.68) and (350.32,103.86) .. (350.32,125.06) .. controls (350.32,146.26) and (332.94,163.45) .. (311.5,163.45) .. controls (290.06,163.45) and (272.68,146.26) .. (272.68,125.06)(265,125.06) .. controls (265,99.62) and (285.82,79) .. (311.5,79) .. controls (337.18,79) and (358,99.62) .. (358,125.06) .. controls (358,150.5) and (337.18,171.13) .. (311.5,171.13) .. controls (285.82,171.13) and (265,150.5) .. (265,125.06) ; \draw  [fill={rgb, 255:red, 0; green, 0; blue, 0 }  ,fill opacity=1 ] (310.8,159) -- (322,167) -- (310.8,175) -- (313.71,167) -- cycle ; \draw  [fill={rgb, 255:red, 0; green, 0; blue, 0 }  ,fill opacity=1 ] (320,90.13) -- (307,83.06) -- (320,76) -- (316.49,83.06) -- cycle ;
\draw  [fill={rgb, 255:red, 128; green, 128; blue, 128 }  ,fill opacity=1 ] (264,123.75) .. controls (264,126.37) and (266.24,128.5) .. (269,128.5) .. controls (271.76,128.5) and (274,126.37) .. (274,123.75) .. controls (274,121.13) and (271.76,119) .. (269,119) .. controls (266.24,119) and (264,121.13) .. (264,123.75) -- cycle ;
\end{tikzpicture} }}$ represent the electron and ion pair-bubble diagrams, respectively,
i.e., $\Pi_{e}$ and $\Pi_{N}$ at the RPA level (see Eq.~(\ref{eq:polarizability})),
and the double wavy line \begin{align}
\label{iron_goulomb}
&\vcenter{\hbox{\begin{tikzpicture}[x=0.25pt,y=0.25pt,yscale=-1,xscale=1]
\draw  [line width=0.75]  (273.96,112.93) .. controls (275.85,115.18) and (277.66,117.33) .. (279.76,117.33) .. controls (281.86,117.33) and (283.67,115.18) .. (285.57,112.93) .. controls (287.46,110.68) and (289.27,108.53) .. (291.37,108.53) .. controls (293.47,108.53) and (295.28,110.68) .. (297.17,112.93) .. controls (299.07,115.18) and (300.88,117.33) .. (302.98,117.33) .. controls (305.08,117.33) and (306.89,115.18) .. (308.78,112.93) .. controls (310.67,110.68) and (312.48,108.53) .. (314.58,108.53) .. controls (316.69,108.53) and (318.5,110.68) .. (320.39,112.93) .. controls (322.28,115.18) and (324.09,117.33) .. (326.19,117.33) .. controls (328.29,117.33) and (330.1,115.18) .. (332,112.93) .. controls (333.89,110.68) and (335.7,108.53) .. (337.8,108.53) .. controls (339.9,108.53) and (341.71,110.68) .. (343.6,112.93) .. controls (345.5,115.18) and (347.31,117.33) .. (349.41,117.33) .. controls (351.51,117.33) and (353.32,115.18) .. (355.21,112.93) ;
\draw  [line width=0.75]  (274.75,117.32) .. controls (276.64,119.57) and (278.45,121.72) .. (280.55,121.72) .. controls (282.65,121.72) and (284.46,119.57) .. (286.35,117.32) .. controls (288.25,115.07) and (290.06,112.92) .. (292.16,112.92) .. controls (294.26,112.92) and (296.07,115.07) .. (297.96,117.32) .. controls (299.86,119.57) and (301.67,121.72) .. (303.77,121.72) .. controls (305.87,121.72) and (307.68,119.57) .. (309.57,117.32) .. controls (311.46,115.07) and (313.27,112.92) .. (315.37,112.92) .. controls (317.47,112.92) and (319.28,115.07) .. (321.18,117.32) .. controls (323.07,119.57) and (324.88,121.72) .. (326.98,121.72) .. controls (329.08,121.72) and (330.89,119.57) .. (332.78,117.32) .. controls (334.68,115.07) and (336.49,112.92) .. (338.59,112.92) .. controls (340.69,112.92) and (342.5,115.07) .. (344.39,117.32) .. controls (346.29,119.57) and (348.1,121.72) .. (350.2,121.72) .. controls (352.3,121.72) and (354.11,119.57) .. (356,117.32) ;
\end{tikzpicture}}}=\vcenter{\hbox{
\begin{tikzpicture}[x=0.25pt,y=0.25pt,yscale=-1,xscale=1]
\draw  [line width=0.75]  (253.96,115.93) .. controls (255.85,118.18) and (257.66,120.33) .. (259.76,120.33) .. controls (261.86,120.33) and (263.67,118.18) .. (265.57,115.93) .. controls (267.46,113.68) and (269.27,111.53) .. (271.37,111.53) .. controls (273.47,111.53) and (275.28,113.68) .. (277.17,115.93) .. controls (279.07,118.18) and (280.88,120.33) .. (282.98,120.33) .. controls (285.08,120.33) and (286.89,118.18) .. (288.78,115.93) .. controls (290.67,113.68) and (292.48,111.53) .. (294.58,111.53) .. controls (296.69,111.53) and (298.5,113.68) .. (300.39,115.93) .. controls (302.28,118.18) and (304.09,120.33) .. (306.19,120.33) .. controls (308.29,120.33) and (310.1,118.18) .. (312,115.93) .. controls (313.89,113.68) and (315.7,111.53) .. (317.8,111.53) .. controls (319.9,111.53) and (321.71,113.68) .. (323.6,115.93) .. controls (325.5,118.18) and (327.31,120.33) .. (329.41,120.33) .. controls (331.51,120.33) and (333.32,118.18) .. (335.21,115.93) ;
\end{tikzpicture}}}+\vcenter{\hbox{
\begin{tikzpicture}[x=0.25pt,y=0.25pt,yscale=-1,xscale=1]
\draw  [line width=0.75]  (165,118.66) .. controls (166.89,121.05) and (168.7,123.33) .. (170.8,123.33) .. controls (172.9,123.33) and (174.71,121.05) .. (176.61,118.66) .. controls (178.5,116.27) and (180.31,114) .. (182.41,114) .. controls (184.51,114) and (186.32,116.27) .. (188.22,118.66) .. controls (190.11,121.05) and (191.92,123.33) .. (194.02,123.33) .. controls (196.12,123.33) and (197.93,121.05) .. (199.82,118.66) .. controls (201.72,116.27) and (203.53,114) .. (205.63,114) .. controls (207.73,114) and (209.54,116.27) .. (211.43,118.66) .. controls (213.32,121.05) and (215.13,123.33) .. (217.23,123.33) .. controls (219.33,123.33) and (221.14,121.05) .. (223.04,118.66) .. controls (224.74,116.52) and (226.37,114.46) .. (228.21,114.07) ;
\draw  [fill={rgb, 255:red, 128; green, 128; blue, 128 }  ,fill opacity=1 ] (314,118.75) .. controls (314,121.37) and (316.24,123.5) .. (319,123.5) .. controls (321.76,123.5) and (324,121.37) .. (324,118.75) .. controls (324,116.13) and (321.76,114) .. (319,114) .. controls (316.24,114) and (314,116.13) .. (314,118.75) -- cycle ; 
\draw   (235.68,120.06) .. controls (235.68,98.86) and (253.06,81.68) .. (274.5,81.68) .. controls (295.94,81.68) and (313.32,98.86) .. (313.32,120.06) .. controls (313.32,141.26) and (295.94,158.45) .. (274.5,158.45) .. controls (253.06,158.45) and (235.68,141.26) .. (235.68,120.06)(228,120.06) .. controls (228,94.62) and (248.82,74) .. (274.5,74) .. controls (300.18,74) and (321,94.62) .. (321,120.06) .. controls (321,145.5) and (300.18,166.13) .. (274.5,166.13) .. controls (248.82,166.13) and (228,145.5) .. (228,120.06) ; \draw  [fill={rgb, 255:red, 0; green, 0; blue, 0 }  ,fill opacity=1 ] (273.8,154) -- (285,162) -- (273.8,170) -- (276.71,162) -- cycle ; \draw  [fill={rgb, 255:red, 0; green, 0; blue, 0 }  ,fill opacity=1 ] (283,85.13) -- (270,78.06) -- (283,71) -- (279.49,78.06) -- cycle ;
\draw  [fill={rgb, 255:red, 128; green, 128; blue, 128 }  ,fill opacity=1 ] (227,118.75) .. controls (227,121.37) and (229.24,123.5) .. (232,123.5) .. controls (234.76,123.5) and (237,121.37) .. (237,118.75) .. controls (237,116.13) and (234.76,114) .. (232,114) .. controls (229.24,114) and (227,116.13) .. (227,118.75) -- cycle ;
\draw  [line width=0.75]  (324,118.66) .. controls (325.89,121.05) and (327.7,123.33) .. (329.8,123.33) .. controls (331.9,123.33) and (333.71,121.05) .. (335.61,118.66) .. controls (337.5,116.27) and (339.31,114) .. (341.41,114) .. controls (343.51,114) and (345.32,116.27) .. (347.22,118.66) .. controls (349.11,121.05) and (350.92,123.33) .. (353.02,123.33) .. controls (355.12,123.33) and (356.93,121.05) .. (358.82,118.66) .. controls (360.72,116.27) and (362.53,114) .. (364.63,114) .. controls (366.73,114) and (368.54,116.27) .. (370.43,118.66) .. controls (372.32,121.05) and (374.13,123.33) .. (376.23,123.33) .. controls (378.33,123.33) and (380.14,121.05) .. (382.04,118.66) .. controls (383.74,116.52) and (385.37,114.46) .. (387.21,114.07) ;
\end{tikzpicture}}}+ \vcenter{\hbox{\begin{tikzpicture}[x=0.25pt,y=0.25pt,yscale=-1,xscale=1]
\draw  [line width=0.75]  (165,118.66) .. controls (166.89,121.05) and (168.7,123.33) .. (170.8,123.33) .. controls (172.9,123.33) and (174.71,121.05) .. (176.61,118.66) .. controls (178.5,116.27) and (180.31,114) .. (182.41,114) .. controls (184.51,114) and (186.32,116.27) .. (188.22,118.66) .. controls (190.11,121.05) and (191.92,123.33) .. (194.02,123.33) .. controls (196.12,123.33) and (197.93,121.05) .. (199.82,118.66) .. controls (201.72,116.27) and (203.53,114) .. (205.63,114) .. controls (207.73,114) and (209.54,116.27) .. (211.43,118.66) .. controls (213.32,121.05) and (215.13,123.33) .. (217.23,123.33) .. controls (219.33,123.33) and (221.14,121.05) .. (223.04,118.66) .. controls (224.74,116.52) and (226.37,114.46) .. (228.21,114.07) ;
\draw  [fill={rgb, 255:red, 128; green, 128; blue, 128 }  ,fill opacity=1 ] (314,118.75) .. controls (314,121.37) and (316.24,123.5) .. (319,123.5) .. controls (321.76,123.5) and (324,121.37) .. (324,118.75) .. controls (324,116.13) and (321.76,114) .. (319,114) .. controls (316.24,114) and (314,116.13) .. (314,118.75) -- cycle ; 
\draw   (235.68,120.06) .. controls (235.68,98.86) and (253.06,81.68) .. (274.5,81.68) .. controls (295.94,81.68) and (313.32,98.86) .. (313.32,120.06) .. controls (313.32,141.26) and (295.94,158.45) .. (274.5,158.45) .. controls (253.06,158.45) and (235.68,141.26) .. (235.68,120.06)(228,120.06) .. controls (228,94.62) and (248.82,74) .. (274.5,74) .. controls (300.18,74) and (321,94.62) .. (321,120.06) .. controls (321,145.5) and (300.18,166.13) .. (274.5,166.13) .. controls (248.82,166.13) and (228,145.5) .. (228,120.06) ; \draw  [fill={rgb, 255:red, 0; green, 0; blue, 0 }  ,fill opacity=1 ] (273.8,154) -- (285,162) -- (273.8,170) -- (276.71,162) -- cycle ; \draw  [fill={rgb, 255:red, 0; green, 0; blue, 0 }  ,fill opacity=1 ] (283,85.13) -- (270,78.06) -- (283,71) -- (279.49,78.06) -- cycle ;
\draw  [fill={rgb, 255:red, 128; green, 128; blue, 128 }  ,fill opacity=1 ] (227,118.75) .. controls (227,121.37) and (229.24,123.5) .. (232,123.5) .. controls (234.76,123.5) and (237,121.37) .. (237,118.75) .. controls (237,116.13) and (234.76,114) .. (232,114) .. controls (229.24,114) and (227,116.13) .. (227,118.75) -- cycle ;
\draw  [line width=0.75]  (324,118.66) .. controls (325.89,121.05) and (327.7,123.33) .. (329.8,123.33) .. controls (331.9,123.33) and (333.71,121.05) .. (335.61,118.66) .. controls (337.5,116.27) and (339.31,114) .. (341.41,114) .. controls (343.51,114) and (345.32,116.27) .. (347.22,118.66) .. controls (349.11,121.05) and (350.92,123.33) .. (353.02,123.33) .. controls (355.12,123.33) and (356.93,121.05) .. (358.82,118.66) .. controls (360.72,116.27) and (362.53,114) .. (364.63,114) .. controls (366.73,114) and (368.54,116.27) .. (370.43,118.66) .. controls (372.32,121.05) and (374.13,123.33) .. (376.23,123.33) .. controls (378.33,123.33) and (380.14,121.05) .. (382.04,118.66) .. controls (383.74,116.52) and (385.37,114.46) .. (387.21,114.07) ;
\draw  [fill={rgb, 255:red, 128; green, 128; blue, 128 }  ,fill opacity=1 ] (474,116.75) .. controls (474,119.37) and (476.24,121.5) .. (479,121.5) .. controls (481.76,121.5) and (484,119.37) .. (484,116.75) .. controls (484,114.13) and (481.76,112) .. (479,112) .. controls (476.24,112) and (474,114.13) .. (474,116.75) -- cycle ; 
\draw   (395.68,118.06) .. controls (395.68,96.86) and (413.06,79.68) .. (434.5,79.68) .. controls (455.94,79.68) and (473.32,96.86) .. (473.32,118.06) .. controls (473.32,139.26) and (455.94,156.45) .. (434.5,156.45) .. controls (413.06,156.45) and (395.68,139.26) .. (395.68,118.06)(388,118.06) .. controls (388,92.62) and (408.82,72) .. (434.5,72) .. controls (460.18,72) and (481,92.62) .. (481,118.06) .. controls (481,143.5) and (460.18,164.13) .. (434.5,164.13) .. controls (408.82,164.13) and (388,143.5) .. (388,118.06) ; \draw  [fill={rgb, 255:red, 0; green, 0; blue, 0 }  ,fill opacity=1 ] (433.8,152) -- (445,160) -- (433.8,168) -- (436.71,160) -- cycle ; \draw  [fill={rgb, 255:red, 0; green, 0; blue, 0 }  ,fill opacity=1 ] (443,83.13) -- (430,76.06) -- (443,69) -- (439.49,76.06) -- cycle ;
\draw  [fill={rgb, 255:red, 128; green, 128; blue, 128 }  ,fill opacity=1 ] (387,116.75) .. controls (387,119.37) and (389.24,121.5) .. (392,121.5) .. controls (394.76,121.5) and (397,119.37) .. (397,116.75) .. controls (397,114.13) and (394.76,112) .. (392,112) .. controls (389.24,112) and (387,114.13) .. (387,116.75) -- cycle ;
\draw  [line width=0.75]  (484,115.66) .. controls (485.89,118.05) and (487.7,120.33) .. (489.8,120.33) .. controls (491.9,120.33) and (493.71,118.05) .. (495.61,115.66) .. controls (497.5,113.27) and (499.31,111) .. (501.41,111) .. controls (503.51,111) and (505.32,113.27) .. (507.22,115.66) .. controls (509.11,118.05) and (510.92,120.33) .. (513.02,120.33) .. controls (515.12,120.33) and (516.93,118.05) .. (518.82,115.66) .. controls (520.72,113.27) and (522.53,111) .. (524.63,111) .. controls (526.73,111) and (528.54,113.27) .. (530.43,115.66) .. controls (532.32,118.05) and (534.13,120.33) .. (536.23,120.33) .. controls (538.33,120.33) and (540.14,118.05) .. (542.04,115.66) .. controls (543.74,113.52) and (545.37,111.46) .. (547.21,111.07) ;
\end{tikzpicture}}}+\cdots\nonumber    \\
\nonumber    \\
&=V_{e}+V_{e}\left(-Z\right)\Pi_{N}V_{e}\left(-Z\right)\nonumber \\
&\qquad+V_{e}\left(-Z\right)\Pi_{N}V_{e}\left(-Z\right)^{2}\Pi_{N}V_{e}\left(-Z\right)+\cdots \nonumber    \\
\nonumber    \\
&=\frac{V_{e}}{1-Z^{2}~V_{e}~\Pi_{N}}
\end{align}represents the electron Coulomb interaction screened by the ions (the
single wavy line corresponds to the Coulomb interaction $V_{e}\left(Q\right)$).
Then Eq.~(\ref{mixing}) can be explicitly written as
\begin{eqnarray}
\chi_{\hat{\rho}_{e}\hat{\rho}_{N}}^{\mathrm{r}} & = & \frac{\Pi_{e}\left(-Z\right)V_{e}\Pi_{N}}{1-V_{e}\,\Pi_{e}-Z^{2}\,V_{e}\,{\displaystyle \Pi_{N}}}.
\end{eqnarray}
\begin{figure}
\begin{centering}
\includegraphics[scale=0.35]{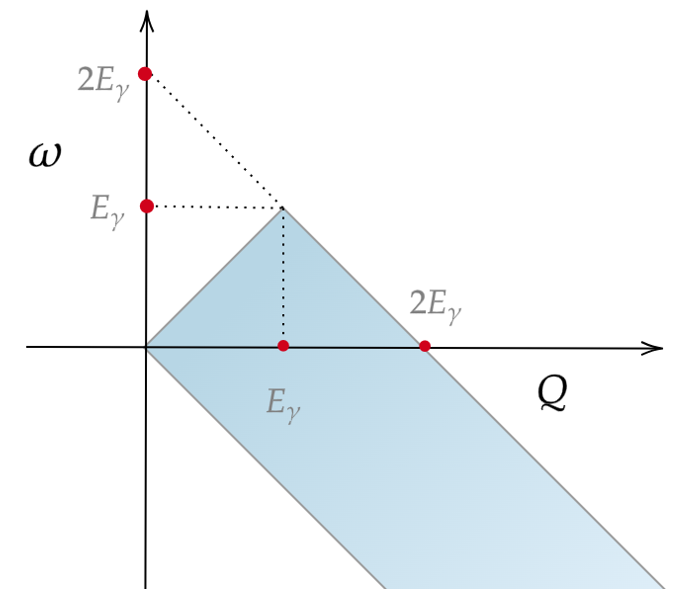}
\par\end{centering}
\caption{\label{fig:InteralArea}The effective integral area relevant for the
Primakoff event rate of Eq.~(\ref{eq:DetailedGamma}). See the text
for details.}
\end{figure}

The above discussion can be extended to obtain the following retarded
correlation functions $\chi_{\hat{\rho}_{N}\hat{\rho}_{e}}^{\mathrm{r}}$
and $\chi_{\hat{\rho}_{N}\hat{\rho}_{N}}^{\mathrm{r}}$ such that,
\begin{eqnarray}
\chi_{\hat{\rho}_{N}\hat{\rho}_{e}}^{\mathrm{r}} & = & \chi_{\hat{\rho}_{e}\hat{\rho}_{N}}^{\mathrm{r}},
\end{eqnarray}
and
\begin{eqnarray}
\chi_{\hat{\rho}_{N}\hat{\rho}_{N}}^{\mathrm{r}} & = & \frac{\left(1-V_{e}\Pi_{e}\right)\Pi_{N}}{1-V_{e}\,\Pi_{e}-V_{e}\,{\displaystyle Z^{2}\,\Pi_{N}}}.
\end{eqnarray}
In addition, in Ref. \citep{Liang:2023ira}, we have already derived
\begin{eqnarray}
\chi_{\hat{\rho}_{e}\hat{\rho}_{e}}^{\mathrm{r}} & = & \frac{\Pi_{e}\left(1-Z^{2}V_{e}\Pi_{N}\right)}{1-V_{e}\,\Pi_{e}-Z^{2}\,V_{e}\,{\displaystyle \Pi_{N}}}.
\end{eqnarray}
Thus, by combining all these terms it is straightforward to verify
\begin{align}
 & \chi_{\hat{\rho}_{e}\hat{\rho}_{e}}^{\mathrm{r}}+\left(-Z\right)\chi_{\hat{\rho}_{e}\hat{\rho}_{N}}^{\mathrm{r}}+\left(-Z\right)\chi_{\hat{\rho}_{N}\hat{\rho}_{e}}^{\mathrm{r}}+\left(-Z\right)^{2}\chi_{\hat{\rho}_{N}\hat{\rho}_{N}}^{\mathrm{r}}\nonumber \\
\nonumber \\
 & =\frac{\Pi_{e}}{1-V_{e}\,\Pi_{e}-Z^{2}\,V_{e}\,{\displaystyle \Pi_{N}}}+\frac{Z^{2}\,\Pi_{N}}{1-V_{e}\,\Pi_{e}-Z^{2}\,V_{e}\,{\displaystyle \Pi_{N}}}\nonumber \\
\end{align}
in Eq.~(\ref{eq:gamma0}). As has been noted in Ref. \citep{Liang:2023ira},
this expression encodes both the thermal movement, and the in-medium
effect of the electrons and ions.

In practical computation of Eq.~(\ref{eq:gamma0}), we first integrate
out the polar angle of $\mathbf{Q}$ with respective to the direction
of the photon momentum $\mathbf{p}_{\gamma}$, which is fixed as the
$z$-axis in the spherical coordinate system. We then take a variable
transformation from $\cos\theta_{\mathbf{Q}\mathbf{p}_{\gamma}}$
to the variable $E_{a}=\sqrt{\left|\mathbf{p}_{\gamma}-\mathbf{Q}\right|^{2}+m_{a}^{2}}$,
along with the corresponding Jacobian
\begin{eqnarray}
\left|\frac{\mathrm{d}\cos\theta_{\mathbf{Q}\mathbf{p}_{\gamma}}}{\mathrm{d}E_{a}}\right| & = & \frac{p_{\gamma}Q}{E_{a}}.
\end{eqnarray}
With this change of variable, the term proportional to $\sin^{2}\theta_{\mathbf{Q}\mathbf{p}_{\gamma}}$
in Eq.~(\ref{eq:gamma0}), i.e., $\left|\mathbf{p}_{\gamma}\times\mathbf{Q}\right|^{2}$
can be rewritten as $p_{\gamma}^{2}Q^{2}$$-\left[\left(E_{a}^{2}-m_{a}^{2}-p_{\gamma}^{2}-Q^{2}\right)^{2}/4\right]$,
and thus Eq.~(\ref{eq:gamma0}) is further expressed as \begin{widetext}
\begin{eqnarray}
\Gamma\left(\mathbf{p}_{\gamma}\right) & = & \int\mathrm{d}\omega\int\frac{\mathrm{d}^{3}Q}{\left(2\pi\right)^{3}}\frac{4\pi\alpha}{Q^{4}}\frac{g_{a\gamma}^{2}\,\left|\mathbf{p}_{\gamma}\times\mathbf{Q}\right|^{2}}{8\,E_{\gamma}\,\sqrt{\left|\mathbf{p}_{\gamma}-\mathbf{Q}\right|^{2}+m_{a}^{2}}}\frac{\left(-2\right)}{1-e^{-\omega/T_{\odot}}}\left[\frac{\mathrm{Im}\left(\Pi_{e}\right)}{\left|1-V_{e}\,\Pi_{e}-V_{e}\,{\displaystyle Z^{2}\,\Pi_{N}}\right|^{2}}+\frac{Z^{2}\,\mathrm{Im}\left(\Pi_{N}\right)}{\left|1-V_{e}\,\Pi_{e}-V_{e}\,{\displaystyle Z^{2}\,\Pi_{N}}\right|^{2}}\right]\nonumber \\
 &  & \times\delta\left(\sqrt{\left|\mathbf{p}_{\gamma}-\mathbf{Q}\right|^{2}+m_{a}^{2}}-E_{\gamma}+\omega\right)\nonumber \\
 & = & \int\mathrm{d}\omega\int\frac{Q\,\mathrm{d}Q}{\left(2\pi\right)^{2}}\frac{4\pi\alpha}{Q^{4}}\frac{g_{a\gamma}^{2}}{8\,E_{\gamma}^{2}}\frac{\left(-2\right)}{1-e^{-\omega/T_{\odot}}}\left[\frac{\mathrm{Im}\left(\Pi_{e}\right)}{\left|1-V_{e}\,\Pi_{e}-V_{e}\,{\displaystyle Z^{2}\,\Pi_{N}}\right|^{2}}+\frac{Z^{2}\,\mathrm{Im}\left(\Pi_{N}\right)}{\left|1-V_{e}\,\Pi_{e}-V_{e}\,{\displaystyle Z^{2}\,\Pi_{N}}\right|^{2}}\right]\nonumber \\
 &  & \times\int_{E_{-}}^{E^{+}}\mathrm{d}E_{a}\left[p_{\gamma}^{2}Q^{2}-\frac{\left(E_{a}^{2}-m_{a}^{2}-p_{\gamma}^{2}-Q^{2}\right)^{2}}{4}\right]\delta\left(E_{a}-E_{\gamma}+\omega\right)\nonumber \\
\nonumber \\
 & = & \int\mathrm{d}\omega\int\frac{Q\,\mathrm{d}Q}{\left(2\pi\right)^{2}}\frac{4\pi\alpha}{Q^{4}}\frac{g_{a\gamma}^{2}}{8\,E_{\gamma}^{2}}\frac{\left(-2\right)}{1-e^{-\omega/T_{\odot}}}\left[\frac{\mathrm{Im}\left(\Pi_{e}\right)}{\left|1-V_{e}\,\Pi_{e}-V_{e}\,{\displaystyle Z^{2}\,\Pi_{N}}\right|^{2}}+\frac{Z^{2}\,\mathrm{Im}\left(\Pi_{N}\right)}{\left|1-V_{e}\,\Pi_{e}-V_{e}\,{\displaystyle Z^{2}\,\Pi_{N}}\right|^{2}}\right]\nonumber \\
 &  & \times\left[p_{\gamma}^{2}Q^{2}-\frac{\left(\omega^{2}-2\omega p_{\gamma}-Q^{2}\right)^{2}}{4}\right]\Theta\left(Q+\omega\right)\cdot\left[\Theta\left(Q-\omega\right)\Theta\left(p_{\gamma}-Q\right)+\Theta\left(2E_{\gamma}-Q-\omega\right)\Theta\left(Q-p_{\gamma}\right)\right],\label{eq:DetailedGamma}
\end{eqnarray}
where $E_{\pm}=\sqrt{\left(p_{\gamma}\pm Q\right)^{2}+m_{a}^{2}}$.
In the last line, we take $m_{a}\rightarrow0$, and $\Theta$ is the
Heaviside step function. The integral area on the $Q$-$\omega$ plane
corresponding to these step functions is shown in Fig.~\ref{fig:InteralArea}
for illustration. In the last step, we generalize the above expression
to the multiple atom species in the Sun (as the sum over isotopes
in the square brackets in Eq.~(\ref{eq:Primakoff rate})).\end{widetext}

\twocolumngrid
\renewcommand{\baselinestretch}{1.1}

\bibliographystyle{JHEP1}
\addcontentsline{toc}{section}{\refname}\bibliography{Primakoff_Sun.bbl}

\end{document}